\documentclass[tradiabstract]{aa}

\usepackage{graphicx}
\usepackage{txfonts}

\usepackage{epsf}
\usepackage{rotating}
\usepackage{graphics}

\newcommand{\emcee}{{\tt emcee}}

\def \1s{$1\,\sigma$}

\def \t0{T$_0$}

\def\m2s2{\hbox{\,m$^{2}$\,s$^{-2}$}} 
     
\def\vsini{\hbox{$v$\,sin\,$i_*$}}  
  
\def\Msun{\hbox{$\mathrm{M}_{\odot}$}}       
\def\Rsun{\hbox{$\mathrm{R}_{\odot}$}}
\def\Mjup{\hbox{$\mathrm{M}_{\rm Jup}$}}
\def\Rjup{\hbox{$\mathrm{R}_{\rm Jup}$}}

\usepackage{amstext}
\usepackage{color}

\begin{document}

   \title{Discovery and characterization of the exoplanets WASP-148b and c}
   \subtitle{A transiting system with two interacting giant planets\thanks{The full 
   version of the SOPHIE measurements (Table~\ref{table_rv})
is only available in electronic form at the CDS via anonymous ftp to 
cdsarc.u-strasbg.fr (130.79.128.5)
or via http://cdsweb.u-strasbg.fr/cgi-bin/qcat?J/A+A/TBC.}}

   \author{
G.~H\'ebrard\inst{1,2}
\and
R.\,F. D\'{\i}az\inst{3,4,5} 
\and
A.\,C.\,M.~Correia\inst{6,7} 
\and
A.~Collier~Cameron\inst{8} 
\and
J.~Laskar\inst{7} 
\and
D.~Pollacco\inst{9,10} 
\and
J.-M.~Almenara\inst{11}
\and
D.\,R.~Anderson\inst{9,10,12} 
\and
S.\,C.\,C.~Barros\inst{13}      
\and
I.~Boisse\inst{14} 
\and
A.\,S.~Bonomo\inst{15}
\and
F.~Bouchy\inst{16} 
\and
G.~Bou\'e\inst{7} 
\and
P.~Boumis\inst{17} 
\and
D.\,J.\,A.~Brown\inst{9,10}             
\and
S.~Dalal\inst{1} 
\and
M.~Deleuil\inst{14} 
\and
O.\,D.\,S.~Demangeon\inst{13}   
\and
A.\,P.~Doyle\inst{9,10} 
\and
C.\,A.~Haswell\inst{18}
\and
C.~Hellier\inst{12}     
\and
H.~Osborn\inst{9,10,14,19}
\and
F.~Kiefer\inst{1,20}
\and
U.\,C.~Kolb\inst{18}
\and
K.~Lam\inst{9,10,21} 
\and
A.~Lecavelier des \'Etangs\inst{1}
\and
T.~Lopez\inst{14} 
\and
M.~Martin-Lagarde\inst{1}
\and
P.~Maxted\inst{12}
\and
J.~McCormac\inst{9,10}
\and
L.\,D.~Nielsen\inst{16}
\and
E.~Pall\'e\inst{22,23}                  
\and
J.~Prieto-Arranz\inst{22,23}            
\and
D.~Queloz\inst{16,24}   
\and
A.~Santerne\inst{14} 
\and
B.~Smalley\inst{12} 
\and
O.~Turner\inst{16} 
\and
S.~Udry\inst{16} 
\and
D.~Verilhac\inst{25} 
\and
R.~West\inst{9,10} 
\and
P.\,J.~Wheatley\inst{9,10} 
\and
P.\,A.~Wilson\inst{9,10} 
}

  \offprints{Guillaume H\'ebrard (hebrard@iap.fr)}

   \institute{
Institut d'astrophysique de Paris, UMR7095 CNRS, Universit\'e Pierre \& Marie Curie, 
98bis boulevard Arago, 75014 Paris, France 
\and
Observatoire de Haute-Provence, CNRS, Universit\'e d'Aix-Marseille, 04870 Saint-Michel-l'Observatoire, France
\and
International Center for Advanced Studies (ICAS) and ICIFI (CONICET), ECyT-UNSAM, Campus Miguelete, 
25 de Mayo y Francia, (1650) Buenos Aires, Argentina
\and
Universidad de Buenos Aires, Facultad de Ciencias Exactas y Naturales. Buenos Aires, Argentina
\and
CONICET - Universidad de Buenos Aires. Instituto de Astronom\'ia y F\'isica del Espacio (IAFE),
Buenos Aires, Argentina
\and
CFisUC, Department of Physics, University of Coimbra,  3004-516 Coimbra, Portugal
\and
IMCCE, UMR8028 CNRS, Observatoire de Paris, PSL University, Sorbonne Univ., 
77 av. Denfert-Rochereau, 75014 Paris, France
\and
School of Physics and Astronomy, Physical Science Building, North Haugh, St Andrews, United Kingdom
\and
Centre for Exoplanets and Habitability, University of Warwick, Gibbet Hill Road, 
Coventry CV4 7AL, United Kingdom
\and
Department of Physics, University of Warwick, Gibbet Hill Road, 
Coventry CV4 7AL, United Kingdom
\and
Universit\'e Grenoble Alpes, CNRS, IPAG, 38000 Grenoble, France
\and
Astrophysics Group, Keele University, Staffordshire, ST5 5BG, United Kingdom
\and
Instituto de Astrof{\'\i}sica e Ci\^encias do Espa\c{c}o, Universidade do Porto, CAUP, Rua das Estrelas, 
4150-762 Porto, Portugal
\and
Laboratoire\,d'Astrophysique\,de\,Marseille,\,Univ.\,de\,Provence,\,UMR6110\,CNRS,\,38\,r.\,F.\,Joliot\,Curie,\,13388\,Marseille\,cedex\,13,\,France
\and
INAF, Osservatorio Astrofisico di Torino, via Osservatorio 20, 10025, Pino Torinese, Italy
\and
Observatoire de Gen\`eve,  Universit\'e de Gen\`eve, 51 Chemin des Maillettes, 1290 Sauverny, Switzerland
\and
Institute for Astronomy, Astrophysics, Space Applications and Remote Sensing, National Obs. of Athens, 15236 Penteli, Greece
\and
School of Physical Sciences, The Open University, Milton Keynes, MK7 6AA, United Kingdom
\and
Center for Space and Habitability, University of Bern, Gesellschaftsstrasse 6, 3012 Bern, Switzerland
\and
LESIA, Observatoire de Paris, Universit\'e PSL, CNRS, Sorbonne Universit\'e, Universit\'e de Paris, 
92195 Meudon, France
\and
Center for Astronomy and Astrophysics, Technical University Berlin, 
Hardenbergstr. 36, 10623 Berlin, Germany
\and
Instituto de Astrof\'isica de Canarias (IAC), E-38200 La Laguna, Tenerife, Spain
\and
Deptartamento de Astrof\'isica, Universidad de La Laguna (ULL), E-38206 La Laguna, Tenerife, Spain
\and
Cavendish Laboratory, J J Thomson Avenue, Cambridge CB3 0HE, United Kingdom
\and
Observatoire Hubert-Reeves, 07320 Mars, France
}

   \date{Received 29 April 2020 / Accepted 16 June 2020}
      
\abstract{
We present the discovery and characterization of WASP-148, a new extrasolar system that includes 
at least two giant planets. The host star is a slowly rotating inactive late-G 
dwarf with a $V=12$ magnitude. The planet
WASP-148b is a hot Jupiter of 0.72\,\Rjup\ 
and 0.29\,\Mjup\ that transits its host 
with an orbital period of 8.80~days.
We found the planetary candidate with the SuperWASP photometric survey, then 
characterized it with the SOPHIE spectrograph. 
Our radial velocity measurements subsequently revealed 
 a second planet in the system, WASP-148c, with an orbital period of 34.5~days
and a minimum mass of 0.40\,\Mjup. 
No transits of this outer planet were detected. 
The orbits of both planets are eccentric and fall near the 4:1 mean-motion resonances. 
This configuration is stable on long timescales, but 
induces dynamical interactions 
so that the orbits differ slightly 
from purely Keplerian orbits. In particular, WASP-148b shows transit-timing variations 
of typically 15~minutes,
making it the first interacting system with transit-timing variations that is
detected on ground-based light curves.
We  establish that the mutual inclination of the orbital plane of the two planets 
cannot be higher than~$35^\circ$, and the true mass of WASP-148c is below 0.60~M$_\mathrm{Jup}$.
We present photometric and spectroscopic observations of this system
that cover a time span of ten years. We also provide their 
Keplerian and Newtonian analyses; 
these analyses should be significantly improved through future TESS~observations.
}

\keywords{Planetary systems -- Techniques: radial velocities -- Techniques: photometric -- 
Techniques: spectroscopic -- Stars: individual: WASP-148}

\authorrunning{H\'ebrard et al.}
\titlerunning{Discovery and characterization of the exoplanets WASP-148b and c}

\maketitle

\section{Introduction}
\label{sect_intro}

Extrasolar planets that transit their host stars are especially interesting.
When they are characterized in photometry and spectroscopy, they allow 
numerous studies and the determination of many of their parameters, including their 
radius and mass. About 700 exoplanets have such a double characterization according to the 
exoplanet archives\footnote{exoplanet.eu, exoplanetarchive.ipac.caltech.edu.}.
A few were first discovered from radial-velocity (RV)  surveys, and photometric
follow-up subsequently revealed their transits 
(e.g., Charbonneau et al.~\cite{charbonneau00}, Motalebi et al.~\cite{motalebi15}).
The vast majority were first 
identified from photometric surveys and were then characterized with RV follow-up, however.
Spectroscopic observations of planetary 
candidates revealed by photometry are used to establish 
or reject their planetary nature, in particular, by measuring the mass of the transiting bodies
using the RV method 
(e.g., H\'ebrard et al.~\cite{hebrard14}, Cooke et al.~\cite{cooke20}). 
They are also used to measure the orbital eccentricity and 
obliquity and to characterize the host stars. Long-term RV follow-up could also reveal 
 additional nontransiting planets in the system
(e.g.,~Christiansen~\cite{christiansen17}, Rey et al.~\cite{rey18}). 
Multiplanetary systems like this are particularly interesting for the studies of their dynamics. 

Dynamics can also be studied in multiplanetary systems when transit-timing variations (TTVs) 
are detected. Whereas in a purely Keplerian orbit the epoch of a transit is 
exactly periodic, several gravitational  perturbations 
can produce small deviations of the transit epochs 
with respect to a perfect periodicity. These Newtonian orbits might be caused 
by orbital decay due to tides (e.g., Birkby et al.~\cite{birkby14}) or to gravitational 
interactions between bodies in multiple systems 
(e.g., Holman \&\ Murray~\cite{holman05}, Agol et al.~\cite{agol05}).
In the case of multiplanetary systems, TTVs have larger amplitudes when the orbital 
periods of the planets are nearly commensurable, that is, in or near mean-motion resonances (MMR). 
This makes the TTV analysis a powerful technique
for characterizing such systems, and in particular, for measuring planetary masses and 
eccentricities, or even detecting 
additional perturbing planets (e.g.,~Nesvorny et al.~\cite{nesvorny12}).   

For years, several attempts have been made to detect TTVs of transiting planets with  
ground-based photometry (e.g., D\'{\i}az et al.~\cite{diaz08}, Maciejewski~et al.~\cite{maciejewski10}),
but most of them later were not confirmed
(e.g., Petrucci et al.~\cite{petrucci20}). TTVs indeed are  difficult to identify as transit 
timing strongly depends on the steepest portions of the light curves (the planetary ingress and egress),
which are short-duration events that are easily subject to systematics, especially from the ground. 
The high-quality long-duration light curves of the \textit{Kepler} space telescope finally allowed 
the first detection of TTVs by Holman et al.~(\cite{holman10}) with the famous case of Kepler-9. 
This star is transited by two  giant planets on $\sim19.2$ and $\sim38.9$~days each. This almost 2:1 MMR
causes TTVs with an amplitude of about one day that are clearly detected with~\textit{Kepler}.

\begin{table}[b]
\begin{center}
\caption{Basic data of the planet-host star WASP-148.}
\begin{tabular}{lrr} 
\hline
IDs:\\
Tycho                   &        3083-295-1   &   (1)  \\
2MASS                   &     J16563135+4418095  & (2)  \\
WISE                    &     J165631.33+441809.2  & (2)  \\
Gaia DR1                &       1358355734609272704   & (3) \\
Gaia DR2                        &        1358355738906114816    & (4)  \\
\hline
RA (J2000)              &      16:56:31.340     & (4) \\
DEC (J2000)             &   $+$44:18:09.55      & (4)  \\       
RA proper motion (mas/yr)       &      $-13.477  \pm 0.047$     & (4) \\
DEC proper motion  (mas/yr)     &      $-27.061  \pm 0.046$     & (4) \\
Parallax (mas)          &   $4.030 \pm 0.026 $     & (4)                        \\
Distance (pc)           &   $248.1 \pm 1.6$    & (4)    \\ 
\hline
Magnitudes:\\
$B$                             &       $13.166 \pm 0.006$               &   (5)                 \\
$V$                             &     $12.247 \pm 0.021 $                &   (5)                         \\
$g'$                            &     $12.677 \pm 0.011 $                &   (5)                         \\
$r'$                            &     $12.028 \pm 0.025 $                &   (5)                         \\
$i'$                            &     $11.893 \pm 0.031 $                &   (5)                         \\
$J$                             &     $10.938 \pm 0.024 $                &(2)    \\
$H$                             &     $10.585 \pm 0.018 $                &(2)    \\
$K_s$                   &     $10.506 \pm 0.017 $                &(2)   \\
$W1$                    &     $10.466 \pm 0.022 $                &(2)   \\
$W2$                    &     $10.519 \pm 0.020 $                &(2)   \\
$NUV$                   &       $18.6508 \pm 0.0667 $            &(6) \\
$G$                             &     $12.0845 \pm 0.0003 $      & (4)          \\
$BP$                    &     $12.4894 \pm  0.0018 $      &(4)          \\
$RP$                    &     $11.5403 \pm 0.0010 $        &(4)         \\
\hline  
\multicolumn{3}{l}{ 
References: 
(1) H\o g et al.~(\cite{hog00}); 
(2) Cutri et al.~(\cite{cutri03}),  
}
\\
\multicolumn{3}{l}{ 
Skrutskie et  al.~(\cite{skrutskie06});
(3) Gaia Collaboration~(\cite{gaia16});
}
\\
\multicolumn{3}{l}{ 
(4) Gaia Collaboration~(\cite{gaia18});
(5) Henden et al.~(\cite{henden15});
}
\\
\multicolumn{3}{l}{ 
(6) Bianchi et al.~(\cite{bianchi17}).}
\\
\end{tabular}
\label{startable}
\end{center}
\end{table}

Today, a few dozen exoplanets have been detected or characterized based on  
their TTVs. Most of these detections were made based on \textit{Kepler} photometry. This includes 
multiplanetary transiting systems as well as single-transiting planets showing TTVs that allow the 
detection and characterization of additional nontransiting planets. Notable cases include KOI-142b, 
a $P$\,=\,10.95 d planet showing TTVs of up to one day that are caused by a nontransiting giant planet with 
a period that is twice longer (Nesvorny et al.~\cite{nesvorny13}) and was eventually detected in RVs 
(Barros et al.~\cite{barros14}), or the seven Earth-size planets orbiting the star Trappist-1, which have periods 
between 1.5 and 18.7~days and TTVs up to a few dozen minutes (Gillon et al.~\cite{gillon17}).

Most TTV systems imply Earth- or Neptune-size planets. Hot Jupiters presenting 
TTVs remain rare, which is one of the reasons why TTVs were late to be detected: the first 
researches were mainly attempted on this type of system. 
Only three are confirmed today.
WASP-4b and WASP-12b are hot Jupiters of 1.3 d and 1.1 d periods, respectively; 
both show long-term deviations from a purely periodic orbit
(Bouma et al.~\cite{bouma19}, Southworth et al.~\cite{southworth19}, 
Maciejewski et al.~\cite{maciejewski16}, Patra et al.~\cite{patra17},~\cite{patra20}, 
Yee et al.~\cite{yee20}).
Several scenarios have been proposed to explain these TTVs, 
whose amplitudes are lower than two minutes. They include stellar activity, 
tide-caused orbital decay, or additional companions.
In the case of WASP-4, Bouma et al.~(\cite{bouma20}) recently showed that 
it could be mostly or entirely produced by the line-of-sight acceleration of the system
(see, however, Baluev et al.~\cite{baluev20}).
The third case is WASP-47b, a hot Jupiter on a 1.1 d period showing TTVs of half a minute that are 
explained by two smaller short-period planets (Becker et al.~\cite{becker15},
Weiss et al.~\cite{weiss17}).
 
Here we present the new planetary system \object{WASP-148}, a fourth case of a hot Jupiter 
showing TTVs, 
and the first interacting system with TTVs detected from the ground.
WASP-148b was first identified as a promising 
transiting-planet candidate by the SuperWASP photometric survey with an orbital period of 8.80~days. 
The RV follow-up with the SOPHIE spectrograph established the planetary nature 
of the transiting object and revealed a second outer giant planet. 
WASP-148c orbits the host star with a period of 34.5~days, which is 
near the 4:1 MMR, apparently without transiting it.
The few photometric transits of WASP-148b that have been observed with ground-based telescopes reveal 
significant deviations from a constant orbital period, with TTV amplitudes of about 
a few minutes. They are likely to be due to WASP-148c, at least~partially.

We present the observations of  WASP-148 in Sect.~\ref{sect_observations}, 
determine the properties of the host star in Sect.~\ref{sect_stellar_parameters}, 
and assess the evidence for the presence of the two planets
in Sect.~\ref{sect_evidences_for_planets}.
Section~\ref{sect_keplerian_characterization} 
describes the Keplerian fit to the data, and 
Sect.~\ref{sect_discussion} discusses  dynamic 
analyses of the system before 
we conclude in Sect.~\ref{sect_conclusion}.

\section{Observations and data reduction}
\label{sect_observations}
 
\subsection{Photometric identification with SuperWASP}
\label{sect_wasp_photometry}

Located on La Palma in the Canary Islands, Spain, SuperWASP-North consists of eight 
Canon 200\,mm $f/1.8$ focal lenses coupled to e2v $2048 \times 2048$~pixel CCDs 
with 13.7" pixels and a field of view of  $7.8^{\circ} \times 7.8^{\circ}$, associated 
with a custom-built photometric reduction pipeline (Pollacco et al.~\cite{pollacco06}). 
It observes with a broadband filter (400-700~nm), and
secured thousands of photometric points over several seasons per star.
Periodic signatures of possible planetary transits are identified in these light curves 
using the algorithms presented by Collier Cameron et al.~(\cite{cameron06}). 

\begin{figure*}
\includegraphics{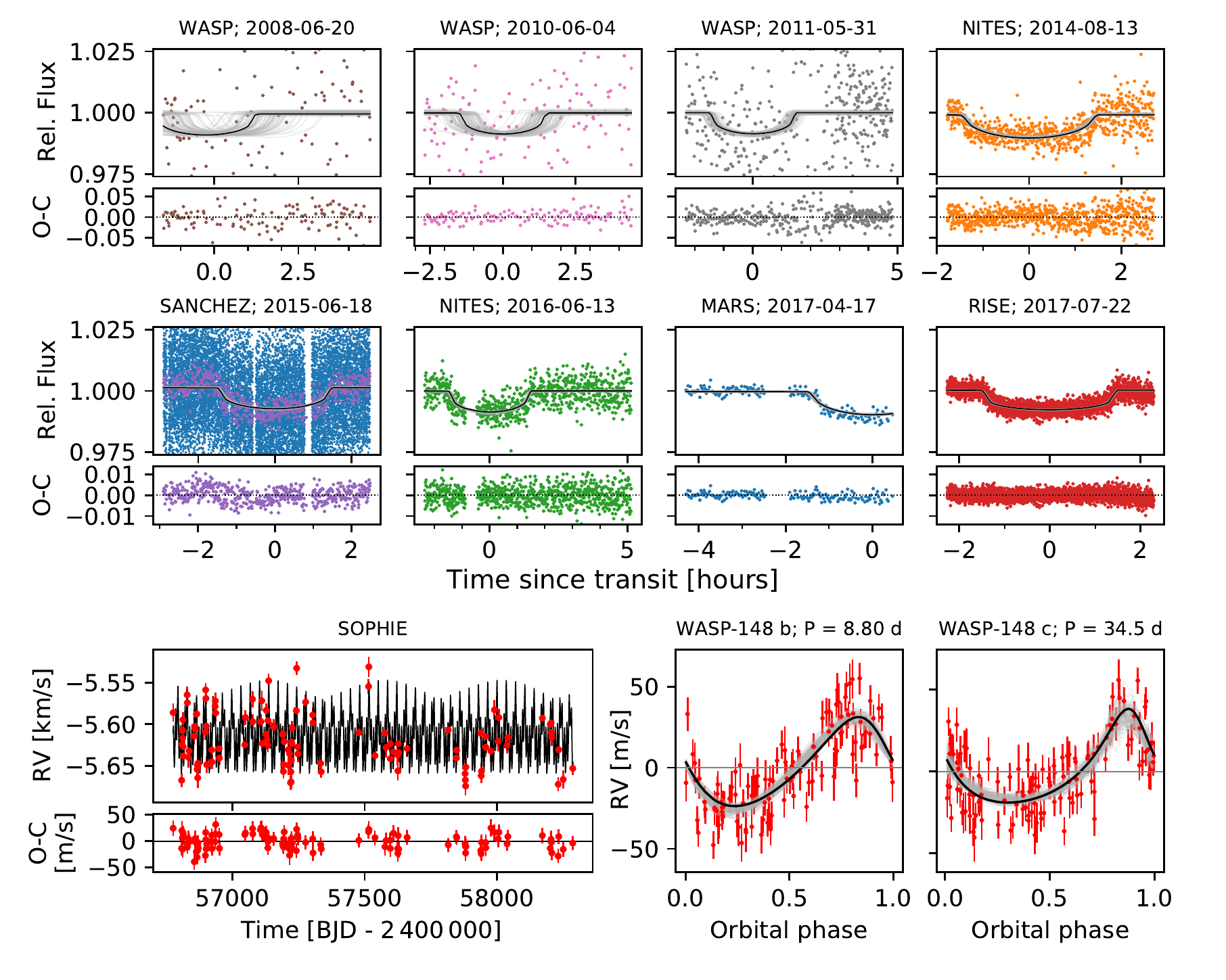}
\caption{Photometric and RV observations of WASP-148, together with their two-planet Keplerian fit 
described in Sect.~\ref{sect_keplerian_characterization} (the parameters are reported in 
Table~\ref{table.params}).
The two upper panels show photometric transit data of WASP-148 acquired by a series of 
observatories (see Table~\ref{table.tcs}). 
The title of each panel shows the observatory name and observation date. 
The relative flux is provided for each transit. Below each panel, the residuals of the MAP 
model are shown. For the transit observed by the S{\'a}nchez 
telescope, we present the full dataset as blue empty circles, and a binned version as purple points.
The lower set of panels shows
the SOPHIE RVs and their 1$\sigma$~error bars
(Table~\ref{table_rv}). 
At left they are plotted as a function of time; 
the residuals to the MAP model are also plotted.
The two  panels at the right are the phase-folded RV curves for WASP-148b ($P=8.80$\,d) 
and WASP-148c ($P=34.5$\,d) after the effect of the other planet is removed.
In  the transit panels and the phase-folded RVs, the solid black curve is the MAP model, 
and the gray thin curves are 100 models drawn randomly from the posterior~distribution
(see Sect.~\ref{sect_keplerian_characterization}).}
\label{fig.lcs}
\end{figure*}

With this facility and procedure, WASP-148 was  identified as the host star of a promising candidate 
for a transiting planet. Three similar planetary-transit-like features were observed on 2008 June 20, 2010 
June 04, and 2011 May 31, with a depth of $\sim0.0070$\,mag, a duration of $\sim3$\,hr, and a possible 
periodicity of $8.80$~days.
The catalog IDs, coordinates, magnitudes, and distance of the star WASP-148 
are reported in Table~\ref{startable}. The SuperWASP light curves are shown 
in the three first plots of Fig.~\ref{fig.lcs};  
the transit-like features are not obvious in these initial data, which illustrates 
how sensitive the candidate detection algorithms should~be.

\subsection{Radial-velocity follow-up with SOPHIE}
\label{sect_RV_meas}

After its identification from SuperWASP photometry, 
we started an RV follow-up of WASP-148 
with the SOPHIE spectrograph at the 1.93 m telescope of the
Observatoire Haute-Provence, France. The goal was first to establish the putative 
planetary nature 
of the transiting candidate, and then in case 
of positive detection, to characterize the planet by measuring notably 
its mass and orbital eccentricity, as well as to search for potential additional bodies in the system.
SOPHIE is a stabilized \'echelle spectrograph dedicated to high-precision RV measurements 
(Perruchot et al.~\cite{perruchot08}, Bouchy et al.~\cite{bouchy09a},~\cite{bouchy13}). 
Here we used its high-efficiency mode with a resolving power $R=40\,000$ and slow 
readout mode to increase the throughput for this faint star.

The first observation season (35 SOPHIE measurements over six months) revealed 
RV variations in phase with the 8.80-day signal detected with SuperWASP, 
no significant variations in the spectral line profiles, and an amplitude of the RV 
variations of about 30\,m/s. 
This showed  that the transiting body is a planet slightly more massive than Saturn, 
designated  WASP-148b hereafter. However, the residuals of the one-planet 
Keplerian fit exhibited a dispersion significantly larger than expected, with 
a possible periodicity.

The second season allowed us to secure a total of 75 measurements over 18 months. The dataset
clearly showed that a second periodic signal was present in the RVs, with a period of about 35~days, 
here again with no significant variations in the spectral line profiles and an RV variation amplitude 
of about 30\,m/s. This indicated the star hosts a second outer planet that is slightly more 
massive than Saturn, hereafter designated WASP-148c

\begin{table}[t]
\caption{SOPHIE measurements of the planet-host star WASP-148}
\begin{center}
\begin{tabular}{cccrrr}
\hline
BJD$_{\rm UTC}$ & RV & $\pm$$1\,\sigma$ & bisect.$^\ast$ & exp. & SNR$^\dagger$ \\
-2\,450\,000 & (km/s) & (km/s) & (km/s)  & (sec) &  \\
\hline
6775.4133 & -5.586 & 0.011 & -0.046 & 1600 & 23.3  \\   
6807.4844 & -5.667 & 0.008 & -0.035 & 1540 & 27.6  \\   
6808.4074 & -5.624 & 0.015 & -0.030 & 1600 & 19.6  \\   
6810.4004 & -5.607 & 0.008 & -0.063 & 1353 & 28.4  \\   
\ldots & \ldots & \ldots  & \ldots  & \ldots  & \ldots \\
\ldots & \ldots & \ldots  & \ldots  & \ldots  & \ldots \\
8231.5736 & -5.672 & 0.009 & -0.018 & 1064 & 28.1  \\   
8233.4430 & -5.631 & 0.009 & -0.015 & 1022 & 27.3  \\   
8250.5540 & -5.666 & 0.011 & -0.010 & 1247 & 21.3  \\   
8286.5702 & -5.653 & 0.008 & -0.027 & 809  & 28.1  \\   
\hline
\multicolumn{6}{l}{The full table is available on line at the CDS.} \\ 
\multicolumn{6}{l}{$\ast$: bisector spans; error bars are twice those of the RVs.} \\ 
\multicolumn{6}{l}{$^\dagger$: S/N per pixel at 550\,nm.} \\
\end{tabular}
\end{center}
\label{table_rv}
\end{table}

The final dataset we present here includes 116 SOPHIE 
measurements secured between April 2014 and June~2018. 
Depending on weather conditions, exposure times ranged from 200 to 2000~seconds with 
a typical value of 1200~seconds. This allowed us to reach a nearly constant signal-to-noise ratio 
S/N\,$=27\pm3$  on each~exposure in order to reduce 
CCD charge transfer inefficiency.
Table~\ref{table_rv} shows the observation log and 
the corresponding barycentric RVs, which were obtained as follows. 
The spectra were extracted using the SOPHIE pipeline (Bouchy et al.~\cite{bouchy09a}) 
and the RVs were measured from the weighted cross correlation with a G2-type 
numerical mask (Baranne et al.~\cite{baranne96}, Pepe et al.~\cite{pepe02}). 
We excluded the 15 bluer SOPHIE spectral orders from the cross correlation as they were particularly noisy.
Spectra were corrected for CCD charge-transfer inefficiency (Bouchy et al.~\cite{bouchy09b}),
and RV error bars were computed from the cross-correlation 
function (CCF) using the method presented by Boisse et al.~(\cite{boisse10}). 
The resulting CCFs have a full width at half maximum (FWHM) of $10.1 \pm 0.1$\,km/s, 
and the contrast represents $\sim35$\,\%\ of the continuum. The final 116-point dataset 
considered here was cleared from measurements secured near the transit of WASP-148b 
to avoid any possible deviation due to the Rossiter-McLaughlin effect, as well as a few 
measurements whose S/N was too low (RVs less accurate than $\pm 15$\,m/s were removed).

The HE mode of SOPHIE is known to present possible instrumental drifts 
(see, e.g., H\'ebrard et al.~\cite{hebrard13}). The causes for these drifts are not 
well understood or identified, but might be due to thermal effects. Following the 
procedure discussed by Santerne et al~(\cite{santerne16}), we used the constant 
star HD\,185144 that was observed on the same nights with SOPHIE in HE mode to correct 
for these potential drifts. The RV dispersion of HD\,185144 is 8.5\,m/s on the 
nights where WASP-148 was observed, with a maximum amplitude of 40~m/s 
observed on a one-month scale in July-August 2015.
This correction allowed a significant improvement of our results.

Following the method described in Pollacco et al.~(\cite{pollacco08}) and
H\'ebrard et al.~(\cite{hebrard08}), for example, we estimated and corrected 
for the moonlight contamination using the second SOPHIE fiber aperture, 
which is targeted on the sky, while the first 
aperture points toward the star. We estimated that 28 spectra of the 116 
were significantly polluted by moonlight. In each of them, the moonlight correction 
ranged from a few to 150\,m/s, with a typical value of about 30\,m/s.
Removing these points does not significantly modify the orbital solution.

Our final SOPHIE dataset thus includes 116 RVs with precisions ranging from 
$6.5$ to $14.9$\,m/s depending on the S/N, with a median value of $9.2$\,m/s. 
They are reported in Table~\ref{table_rv} and displayed in 
the lower panel of Fig.~\ref{fig.lcs}.
The observed 31.0 m/s dispersion is significantly higher than the estimated  
error bars on the measurements, indicating variability.

\begin{table*}[t]
\centering
\caption{Photometric observations, measured jitters, and TTVs of WASP-148b transits.
The four last columns summarize the statistics for the marginal posteriors 
(Sect.~\ref{Sect_Models_and_parametrisation}).
In particular, the TTVs below are obtained using the mean ephemeris derived in  Sect.~\ref{sect.ttvs}, 
and the reported error on the TTV amplitude for each transit corresponding to the error on $T_0$.}
\renewcommand{\footnoterule}{}  
\begin{tabular}{lllccccc}        
\hline             
Instrument      & Band       & Date                     &       Relative num-    &Transit epoch $T_{ \rm 0}$                     &       TTV     &   TTV   & photometric jitter \\
(cf. Fig.~\ref{fig.oc}) &                    &                          &       ber of transit              & [$\rm BJD_{\rm UTC}-2\,450\,000$]             &       [min]           & significance &  [relative flux]  \\
\hline
SuperWASP               & --                 & 2008 Jun 20      &       -377    &       $4638.453 \pm     0.026$          &       $+3 \pm 38$     &       $0.1\, \sigma$  &  $(3.3 \pm 2.5) \times 10^{-3}$  \\     
SuperWASP               & --                 & 2010 Jun 4       &       -296 &       $5351.525 \pm   0.017$          &       $-51 \pm        25$     &       $2.0\, \sigma$ &  $(1.6  \pm 1.3) \times 10^{-3}$  \\  
SuperWASP               & --                 & 2011 May 31      &       -255    &       $5712.5185 \pm 0.0061$             &       $+3.5 \pm 8.6$  &       $0.4\, \sigma$  &  $(4.7 \pm 1.8) \times 10^{-3}$  \\     
Nites           & --                 & 2014 Aug 13      &       -122    &       $6883.4435 \pm 0.0014$             &       $+30 \pm 1.9 $ &        $15.9\, \sigma$  &  $(2.6 \pm 2.0) \times 10^{-4}$   \\  
S{\'a}nchez     & Johnson-R & 2015 Jun 18       &       -87     &       $7191.55670 \pm 0.00082     $       &       $+0.5 \pm       1.0$    &       $0.5\, \sigma$  &  $(3.9 \pm 3.0) \times 10^{-4}$  \\  
Nites           & Johnson-R & 2016 Jun 13       &       -46     &       $7552.5045 \pm 0.0012$             &       $-11.4 \pm      1.5$    &       $7.4\, \sigma$  &  $(3.34 \pm 0.14) \times 10^{-3}$  \\        
Mars            & --                 & 2017 Apr 17      &       -11     &       $7860.6467 \pm 0.0013$             &       $+1.8 \pm       2.2$    &       $0.8\, \sigma$   &  $(9.7 \pm 1.4) \times 10^{-4}$ \\  
Rise                    & og515+kg5 &2017 Jul 22        &       0       &       $7957.48077 \pm 0.00030$    &       $-10.01 \pm     0.41$ & $24.7\, \sigma$ &  $(7.0 \pm 5.4) \times 10^{-5}$  \\    
\hline
\vspace{-0.2cm}
\end{tabular}
\label{table.tcs}  
\end{table*}

\subsection{Additional photometry}
\label{sect_FU_photom}

After the RVs established that the of 8.80-day signal identified with 
SuperWASP was indeed due to a planet, we obtained five additional transit light 
curves with four different larger ground-based telescopes. They allowed 
improved spatial and temporal resolutions as well as more precise time-series photometry during transits. 
The goal was to refine the determination of the parameters derived from photometry.
These observations are briefly described below. Data reductions were standard and 
include bias and flat-field corrections, aperture photometry, comparison stars 
selected to minimize the scatter out of transit, and flux normalization.
The images revealed no contamination on the star.
The logs of the photometric observations are reported in Table~\ref{table.tcs}, 
and the five corresponding transit light curves (four complete transits and one partial 
transit) are plotted in~Fig.~\ref{fig.lcs}; they show obvious transit detections.

Table 3 also shows the measured epochs of each transit measured below 
in Sect.~\ref{sect_keplerian_characterization}. The epochs  exhibit significant deviations 
from the constant ephemerides. We~verified that the times reported by the observers were~correct.
In addition, these telescopes and their clocks have been regularly used for other 
planetary transit studies without showing any timing problems (e.g., 
Hay et al.~\cite{hay16}, Spake et al.~\cite{spake16}, 
Demangeon et al.~\cite{demangeon18}).
We therefore concluded the WASP-148b transit light curves  show indications of~TTVs.

We did not secure dedicated photometric observations to search for any 
possible transit of the outer 34.5-day planet. An inspection of the SuperWASP
light curves did not show any significant signature, but their time coverage is poor.
We can therefore only conclude here  the planet WASP-148c does not show obvious 
signatures of transits.

\subsubsection{Nites}

Two full transits of WASP-148b were observed with 
the Near Infra-red Transiting ExoplanetS (NITES) Telescope: a first transit without filter on 2014 August 13, 
and a second transit with a Johnson-R filter on 2016 June 13.
NITES, located at La Palma in the Canary Islands, Spain, 
is a semirobotic, 0.4-m ($f/10$) Meade LX200GPS Schmidt-Cassegrain telescope 
(McCormac et al.~\cite{mccormac14}). It is
mounted with a Finger Lakes Instrumentation Proline 4710 camera and a $1024\times1024$ 
pixel deep-depleted CCD made by e2v. The telescope has a field of view of 
$11\arcmin\times11\arcmin$ and a pixel scale of $0.66$\arcsec\,pixel$^{-1}$.

\subsubsection{S{\'a}nchez}

A full transit of WASP-148b was observed on 2015 June 18 at the 1.52 m Carlos S{\'a}nchez Telescope 
(TCS) at the Teide Observatory, Tenerife, in the Canary Islands, Spain (Oscoz et al.~\cite{oscoz08}). 
We used the Wide-FastCam camera with a 
Johnson-R filter, and the telescope  was manually guided.
The mean FWHM through the night was 3.30~pixels, and the best aperture for data reduction was 9~pixels.

\subsubsection{Mars}

One partial transit of WASP-148b was observed at Observatoire \textit{Hubert Reeves} in Mars, France, without filter 
on 2017 April 17. A 0.6 m ($f/8$) telescope was used with a $2750\times2200$-pixel 
Atik 460EX camera, and successive exposures of 80 s durations. The observations and their reduction 
were made by amateur astronomers, using the AudeLa and Muniwin softwares.
The telescope has a field of view of $8.68\arcmin\times6.9\arcmin$, and a pixel scale of $0.77$\arcsec\,pixel$^{-1}$. 

\subsubsection{Rise}

A full transit of WASP-148b was observed on 2017 July 22 with a V+R (og515+kg5) filter using the Rise 
instrument mounted on the robotic 2.0 m Liverpool Telescope at La Palma in the Canary Islands, Spain. 
Rise is equipped with a back-illuminated frame-transfer $1024 \times 1024$ pixel CCD. 
The scale is 1.08\arcsec\,pixel$^{-1}$ , and a $2\times2$ binning was used. 
A total of 2720 6 s exposures 
was secured in a row to have a good sampling of the transit. 
This is the best transit light curve of  WASP-148b in the dataset presented~here.

\subsection{Lucky imaging}
\label{sect_luckyimaging}

To further investigate the possibility of stellar contamination of our photometric 
light curves, on 2016 March 9 we carried out a lucky imaging search for additional 
companions around WASP-148 using FastCam (Oscoz et al.~\cite{oscoz08})
on the Carlos S{\'a}nchez Telescope at the Teide Observatory, Tenerife, Spain. 
FastCam has a field of view and pixel scale of $6\arcsec$ and $0.042\arcsec$\,pixel$^{-1}$. The detector is an L3 electron-multiplying CCD (EMCCD) with rapid readout, 
and essentially zero readout noise. We obtained ten data cubes for WASP-148, each 
with 1000 images of 50\,ms exposure time, using no filter. 
The data in each cube were bias-subtracted, aligned, and stacked to increase 
the S/N. 
The resulting data do not allow us to quantify the magnitude 
contrast well, but we can conclude that no visible companion objects were 
found within $6\arcsec$ of the star.
This means that there are no indications here for contamination or~blend.

\section{Stellar properties of WASP-148}
\label{sect_stellar_parameters}

\subsection{Rotation periods}
\label{sect_rot_period}

We used the sine-wave fitting method described in Maxted et al.~(\cite{maxted11})
to search for quasi-periodic modulation in the SuperWASP light curves of WASP-148
caused by the combination of the stellar rotation and magnetic activity, that is,
star spots.  Variability due to star spots is not expected to be coherent on
long timescales as a consequence of the finite lifetime of star spots and
differential rotation in the photosphere. We therefore analyzed each season of data
for WASP-148 separately. We also analyzed the data from each camera that was used to
observe WASP-148 separately so that we could assess the reliability of the
results. Only combinations of cameras and seasons with more than 2000
observations were included in this analysis.  We removed the transit signal
from the data prior to calculating the periodograms by subtracting a simple
transit model from the light curve. We calculated periodograms over 8192
uniformly spaced frequencies from 0 to 0.5 cycle/day. The false-alarm
probability was calculated using the bootstrap Monte Carlo method 
described in Maxted et al.~(\cite{maxted11}).

The results of this analysis are given in Table~\ref{ProtTable}. The best-fit
period for all the light curves obtained with camera 143 are consistent with a
rotation period of about 26 days or its first harmonic. This supposed rotation
period is not detected in the data obtained with camera 141. This may be
because the data from this camera are more strongly affected by instrumental
noise than the data from camera 143. The mean period from the data obtained
with camera 143 is $26.2 \pm 1.3$\,days. 
This is the value we adopted for the stellar rotation period $P_{\rm rot}$.
The amplitude of this long-term modulation is of a few millimagnitudes and therefore
is not visible in the light curves plotted in Fig.~\ref{fig.lcs}, which focus on short-duration transits alone.

\begin{table}
 \caption{Periodogram analysis for long-term sinusoidal 
modulations of the SuperWASP light curves of WASP-148. Observing
dates are  in BJD--2\,450\,000,  $N$ is the number of observations used in the analysis, and
$A$ is the semiamplitude of the best-fit sine wave at the
period $P$ found in the periodogram with a false-alarm probability FAP.
\label{ProtTable}}
 \begin{tabular}{@{}cccccc}
\hline
  \multicolumn{1}{@{}l}{Camera} &
  \multicolumn{1}{c}{Dates} &
  \multicolumn{1}{c}{$N$} &
  \multicolumn{1}{c}{$P$ [d]} &
  \multicolumn{1}{c}{$A$ [mmag]} &
  \multicolumn{1}{c}{FAP}\\
 \noalign{\smallskip}
\hline
141 & 4189\,--\,4316 & 9577 &  39.5 &  5.4 & 0.002  \\
141 & 4553\,--\,4681 & 8173 & 102 &  9.1 & $<0.001$ \\
143 & 4189\,--\,4316 &10104 &  23.1 &  2.3 & 0.004  \\
143 & 4553\,--\,4681 & 8395 &  23.9 &  3.7 & $<0.001$ \\
143 & 4921\,--\,5046 & 8704 &  27.0 &  2.9 & $<0.001$ \\
143 & 5283\,--\,5411 &10408 &  30.8 &  2.3 & $<0.001$ \\
143 & 5648\,--\,5777 &28833 &  13.1 &  2.6 & 0.001  \\
 \noalign{\smallskip}
\hline
 \end{tabular}   
 \end{table}

 \subsection{Spectral characterization}
\label{sect_spectr_charact}
 
The host-star SOPHIE spectra unpolluted by moonlight were RV-corrected and 
averaged to produce a single spectrum. It was used for spectral analysis using 
the methods described in Doyle et al.~(\cite{doyle13}). We used the H$\alpha$ 
line to estimate the effective temperature ($T_{\rm eff}$) and the Na~{\sc i} D and Mg~{\sc i} b 
lines as diagnostics of the surface gravity ($\log g$). The iron abundances $[$Fe/H$]$ were determined 
from equivalent-width measurements of several clean and unblended Fe~{\sc i} lines and are 
given relative to the solar value presented in Asplund et al.~(\cite{asplund09}).
The derived abundance errors include the uncertainties in $T_{\rm eff}$ 
and $\log g$, as well as the scatter that is due to measurement and atomic data uncertainties. 
The projected rotation velocity (\vsini) was determined to be about 2~km/s by fitting the profiles of 
the Fe~{\sc i} lines, after convolving with the SOPHIE instrumental resolution 
and a macroturbulent velocity of $2.8\pm0.7$\,km/s adopted from the calibration of 
Doyle et al.~(\cite{doyle14}).

\begin{figure}[h] 
\begin{center}
\includegraphics[scale=0.35]{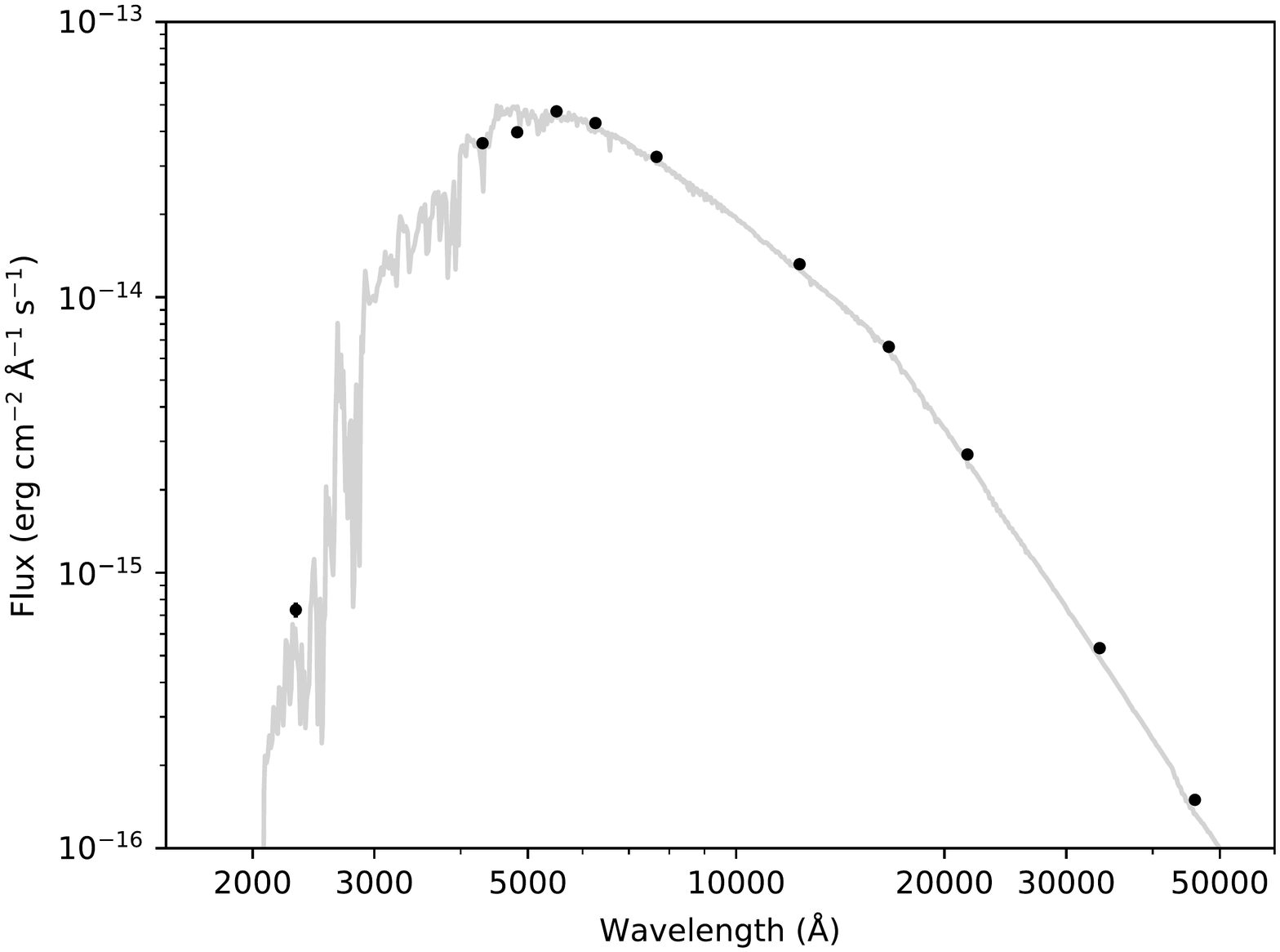}
\caption{Fitted spectral energy distribution of  the star WASP-148 
(gray line) overplotted with measured magnitudes (black circles, 
Table~\ref{startable}).}
\label{fig_SED}
\end{center}
\end{figure}

We obtained $T_{\rm{eff}} = 5460 \pm 130$\,K, log\,$g = 4.40 \pm 0.15$ (cgs),  [Fe/H] = $+0.11 \pm 0.08$, and
$\log$A(Li) $<0.5$. Using the calibration of Torres et al.~(\cite{torres10}), we derived a stellar mass and radius of 
$1.00 \pm 0.08$\,\Msun\ and $1.03 \pm 0.20$\,\Rsun, respectively.
These values agree with 
the values of $T_{\rm{eff}} = 5350 \pm 115$\,K and $R_* = 0.97 \pm 0.04$\,\Rsun\ 
from the Gaia DR2 catalog (Gaia Collaboration~\cite{gaia18}).
As a sanity check, $T_{\rm eff}$ was also obtained from the 
spectral energy distribution (SED). This was obtained using broad-band photometry from 
Table~\ref{startable}, except for the three from Gaia Collaboration~(\cite{gaia18})
in order to be independent of the DR2 $T_{\rm{eff}}$ value.
The photometry was converted into fluxes and the best-fitting 
Kurucz~(\cite{kurucz93}) model flux distribution was determined, which gave a value of 
$T_{\rm eff} = 5430 \pm 140$~K, which again agrees well (see Fig.~\ref{fig_SED}).

The SOPHIE spectra show no chromospheric emission peaks in the Ca\,{\sc ii} H+K lines, 
whereas emission would be a sign of stellar activity.
We computed the activity index $\log{R'_\mathrm{HK}} = -5.09\pm0.11$ from the spectra, 
and the projected rotational velocity \vsini$=4.1 \pm 1.0 $\,km/s
from the parameters of the CCF, both using the calibrations of Boisse et al.~(\cite{boisse10}).
The analysis of spectral lines themselves provided \vsini\,$\sim 2$\,km/s,
therefore we finally adopted the conservative value \vsini$=3 \pm 2 $\,km/s.
This is consistent with the stellar rotation period and radius  reported above,
but the large error bars are not constraining~here.

The $\log{R'_\mathrm{HK}}$ value is consistent with the basal limit corresponding to a 
quiet star. The low RV and photometric jitters found below in 
Sect.~\ref{sect_system_parameters} corroborate this.
The values of $\log{R'_\mathrm{HK}}$ of some hot-Jupiter hosts apparently 
lie below this basal limit, which is inconsistent with our understanding of the 
behavior of late-type stars. This might
be explained by absorption in circumstellar gas that is ablated from the hot planets. This gas will absorb in cores of the Ca\,{\sc ii} H+K lines 
(Haswell et al.~\cite{haswell12},~\cite{haswell19}, Fossati et al.~\cite{fossati13}, 
Staab et al.\cite{staab17}).
This circumstellar absorption phenomenon may operate in the WASP-148 
system, but it has not decreased the $\log{R'_\mathrm{HK}}$ value below the basal limit. 
The intrinsic activity of WASP-148 might consequently be slightly above that 
indicated by $\log{R'_\mathrm{HK}}$ taken at face value, but there is nothing to suggest 
that this effect is significant here as the star is~quiet.
Based on all these analyses, we conclude that WASP-148 is a slowly rotating inactive late-G~dwarf.

\section{Evidence for planets orbiting WASP-148}
\label{sect_evidences_for_planets}

\subsection{Radial velocity periodograms}
\label{sect_RVperiodograms}

\begin{figure}[b!] 
\begin{center}
\includegraphics[scale=0.59]{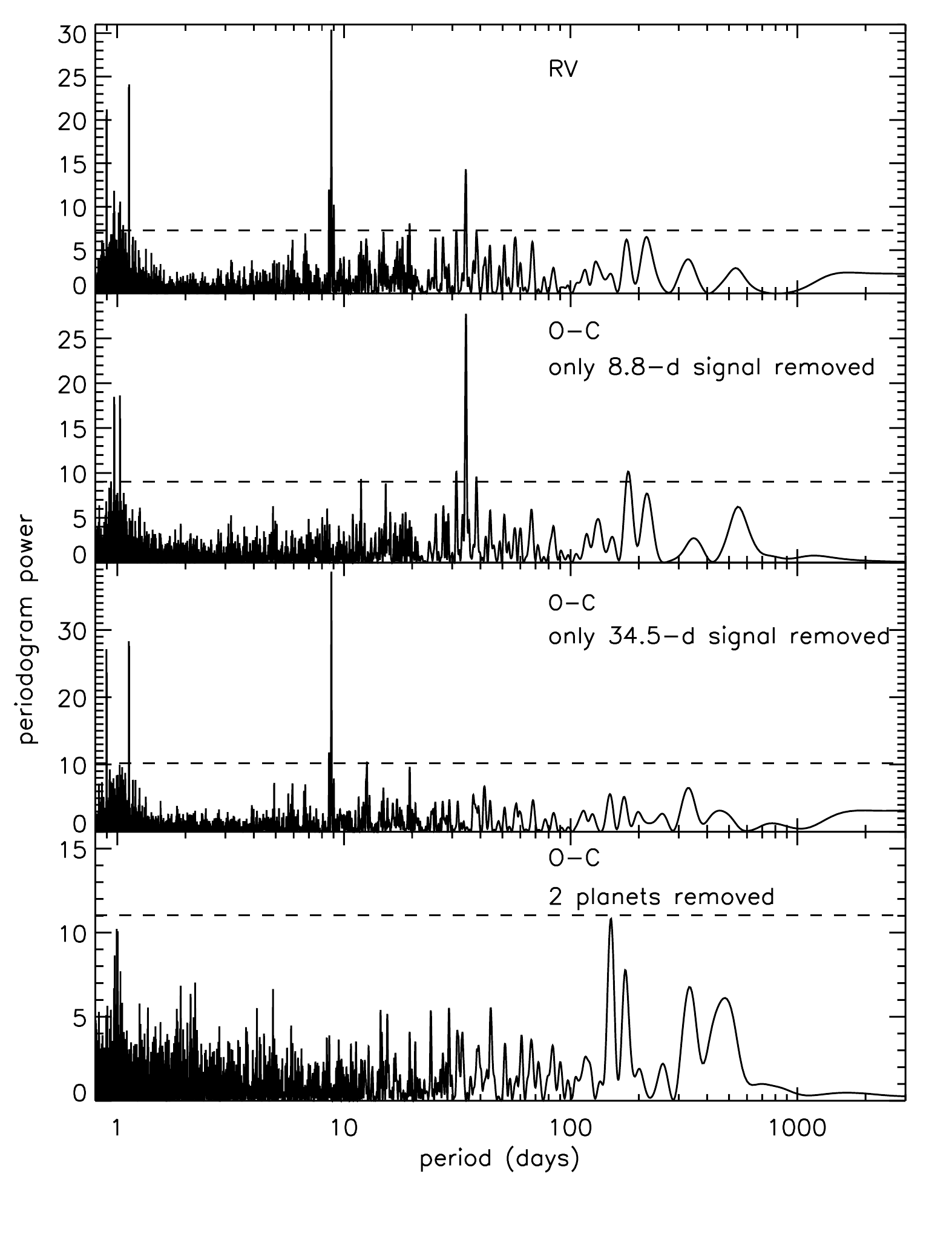}
\caption{Lomb-Scargle periodograms of the SOPHIE RVs of WASP-148. The upper 
panel shows the periodogram computed based on the initial RVs without any fit 
removed. The second and third panels show the periodograms computed based on the residuals 
of the fits including WASP-148b or WASP-148c, respectively. The bottom panel shows the 
periodogram after subtraction of the two-planet fit. 
The horizontal dashed lines correspond to the false-alarm probability of~$1\times10^{-3}$.
}
\label{fig_periodograms_RV}
\end{center}
\end{figure}

In order to study the periodic signals that might be present in our final SOPHIE dataset, 
Fig.~\ref{fig_periodograms_RV} presents their Lomb-Scargle periodograms 
(Press et al.~\cite{press92}) in four different cases, 
as well as the limits corresponding to false-alarm probabilities of~$1\times10^{-3}$.
The upper panel shows the periodogram of the WASP-148  RVs. Two main 
significant peaks are clearly detected at periods $\sim8.80$~days and $\sim34.5$~days,
corresponding to the two planets reported above, together with their weaker aliases 
around one~day.

The second panel shows the periodogram of the RV
residuals to a fit including the inner transiting planet alone.
The standard deviation of the residuals of this fit is 
$\sigma_\mathrm{O-C}=21.3$~m/s, indicating that an additional signal is present.
In this periodogram, the peak at 8.80~days and its aliases 
are no longer visible, as expected. The main remaining peak is the peak at 34.5~days, 
together with the two fainter double-peaks near 0.97 and 1.03\,d that correspond 
to its aliases with one synodic and one sidereal day due to ground-based sampling.
Similarly, the third panel  of Fig.~\ref{fig_periodograms_RV} shows the periodogram of the 
RV residuals to a fit including the outer planet alone (here again with a large 
residuals standard deviation $\sigma_\mathrm{O-C}=25.3$~m/s). The peak at 34.5~days and 
its aliases are removed, and the 8.80-day signal remains together with the two alias double-peaks 
near 0.90 and~1.30\,d.

Finally, the bottom panel of Fig.~\ref{fig_periodograms_RV} shows the periodogram of the 
residuals after the two-planet fit shown in Fig.~\ref{fig.lcs}.
Only lower peaks remain, with false-alarm probabilities below $1\times10^{-3}$.
We note that in addition to the peaks around one day 
(which are due to the aliases of all the detected signals), all the four panels in 
Fig.~\ref{fig_periodograms_RV} show possible peaks at longer periods, in particular 
near 150~days. This might be the signature of a third outer planet. 
It would have an RV semiamplitude of aobut 10~m/s that corresponds to a
sky-projected mass of $\sim0.25\,$\Mjup. However, 
this long-period signal is not strong enough to claim any detection
with the available data. Further 
observations of this star on a longer time baseline are necessary to establish or discard 
the presence of a third planet in this~system.

We also note that none of the periodograms shows any significant power near 26~days, 
as the one seen in SuperWASP photometry and possibly linked to stellar rotation 
(Sect.~\ref{sect_rot_period}). Possible RV signals caused by stellar effects are discussed 
in more detail in the following subsection.

In conclusion, this 
analysis clearly shows that the SOPHIE RVs  include a periodic signal at 8.80~days. This 
corresponds to the signal that was detected in photometry, together with an additional signal at 34.5~days.

\subsection{Validation in planetary nature of the RV signals}

Here we argue that both RV periods are caused by Doppler shifts that are caused by planets orbiting 
WASP-148, and not by spectral line profile variations that are due to stellar activity or blended binaries.
It is well known that 
stellar blended configurations can mimic planetary transits, including undiluted eclipsing binaries with 
low-mass stellar companions or diluted eclipsing binaries  
(e.g.,~Almenara et al.~\cite{almenara09}). 
The astrometric excess noise of WASP-148 measured by Gaia is 0.7~mas, which is
below its detection limit (Kiefer et al.~\cite{kiefer19}), 
thus showing no signatures for contamination or a blend caused by a possible 
massive companion. This agrees with the lack of companion detection in  the 
images obtained for photometry (Sect.~\ref{sect_FU_photom}) as well as in 
our high-resolution imaging (Sect.~\ref{sect_luckyimaging}).
In addition, RVs measured using different stellar masks (F0, K0, or K5) produce 
variations with similar amplitudes to those obtained with the G2 mask, therefore it is unlikely 
that these variations are produced by blend scenarios 
composed of stars of different spectral~types.

Similarly, the measured CCF bisector spans (Table~\ref{table_rv}) quantify possible shape variations 
of the spectral lines. A correlation between the bisector and the 
RV might indeed  be the signature of RV variations induced by blend configurations or stellar activity
(see, e.g., Queloz et al.~\cite{queloz01}, Boisse et al.~\cite{boisse09}). Here, 
bisector spans show a dispersion of 21.9\,m/s, which is 1.4 times smaller than the RV dispersion, 
whereas each bisector span is roughly half as precise as the 
corresponding RV measurement. 
This means that whereas the RV dispersion is caused by the  periodic RV signals, 
the bisector spans show no significant variations.
Moreover, they show no correlations with the RVs. 
The linear correlation parameter is $0.00 \pm 0.07,$ and 
the Pearson and Spearman rank correlation factors 
have low values of -0.01 and -0.06, respectively.
This agrees with the conclusion that the RV~variations are caused by planetary signals 
and not by spectral-line profile changes that are attributable to blends or stellar~activity.

We made the same tests between bisector spans and RV residuals after fitting WASP-148b
alone, WASP-148c alone, or both planets b and c; in none of these cases did we detect any 
correlation. In addition, we made the same tests with the FWHMs of the CCF, which as the bisector spans
quantify the shape of the spectral lines; here we found no correlation between this parameter 
and the RV or their residuals either. Finally, we computed the Lomb-Scargle periodograms of 
bisector spans and FWHMs; they present no significant periodicities, in particular 
at the periods of the two signals seen in RVs.

All of this strengthens the inference that the RV variations are not caused by 
spectral-line profile changes attributable to blends or stellar activity. 
We conclude that they are caused by exoplanets with orbital periods of 
8.80 and 34.5~days.
The detected mutual interactions between the planets provide an additional 
argument for this interpretation (see below in Sect.\ref{sect_fit_ttv}).

\begin{table*}[t!]
\caption{Prior distributions for the Keplerian model parameters.}
\centering
\label{table.priors}
\begin{tabular}{llcc}
\hline\noalign{\smallskip}
&& \multicolumn{1}{c}{\bf WASP-148b} & \multicolumn{1}{c}{\bf WASP-148c}\\
\hline\noalign{\smallskip}
\emph{\hspace{0.3cm} Orbital parameters:} \\
Orbital period, $P$ & [d] & $N(8.8036930, 1.24\times10^{-5})$ & $N(34.5246, 0.18)$\\
\noalign{\smallskip}
Time of inferior conjunction or transit, $T_c$ & [BJD] &$N(2\,457\,957.4807374, 0.00056)$& $--$ \\
\noalign{\smallskip}
Mean longitude at at BJD$\;= 2\,455\,500$, $\lambda_0$ & [deg] & $--$ & $U(0, 360)$\\
\noalign{\smallskip}
RV amplitude, $K_1$ & [m/s] & $U(0, 500)$& $U(0, 100)$ \\
\noalign{\smallskip}
$\sqrt{e} \cos{\omega}$ &  & $U(-0.9, 0.9)$& $U(-0.9, 0.9)$ \\
\noalign{\smallskip}
$\sqrt{e} \sin{\omega}$ &  & $U(-0.9, 0.9)$& $U(-0.9, 0.9)$ \\
\noalign{\smallskip}
Impact parameter, $b$ &  & $U(0, 1)$&$--$ \\
\noalign{\smallskip}
Radius ratio, $R_p/R_*$ &  & $J(0.01, 0.5)$&$--$ \\
\noalign{\smallskip}
Stellar density, $\rho_*$ & [$\rho_\odot$]  & \multicolumn{2}{c}{$tN(1.19, 1.15, 0, 20)$} \\
\noalign{\smallskip}
\hline\noalign{\smallskip}
\emph{\hspace{0.3cm} Linear limb-dakerning coefficients:} \\
Johnson-R &&\multicolumn{2}{c}{$U(0, 1)$} \\
\noalign{\smallskip}
og515+kg5 &&\multicolumn{2}{c}{$U(0, 1)$} \\
\noalign{\smallskip}
Clear &&\multicolumn{2}{c}{$U(0, 1)$} \\
\noalign{\smallskip}
\hline\noalign{\smallskip}
\emph{\hspace{0.3cm} Data parameters:} \\
Timing offset, $\delta T_c$, (all observatories) & [days] & \multicolumn{2}{c}{$N(0, 0.05)$}\\
\noalign{\smallskip}
Flux out of transit, $f_{\mathrm{OOT}_i}$ (all observatories)&& \multicolumn{2}{c}{$N(1, 0.0005)$}\\
\noalign{\smallskip}
Photometric jitter, $\sigma_{LC_i}$ (all observatories)&& \multicolumn{2}{c}{$U(0, 0.08)$}\\
\noalign{\smallskip}
Barycentric systemic RV, $\gamma$ &[km/s]&\multicolumn{2}{c}{$U(4, 6)$}\\
RV jitter, $\sigma_{RV}$ &[m/s]&\multicolumn{2}{c}{$U(0, 120)$}\\
\hline\noalign{\smallskip}
\multicolumn{4}{l}{$U(x_\text{min};  x_\text{max})$: uniform distribution between $x_\text{min}$ and $x_\text{max}$.} \\
\multicolumn{4}{l}{$J(x_\text{min};  x_\text{max})$: Jeffreys (log-flat) distribution between $x_\text{min}$ and $x_\text{max}$.} \\
\multicolumn{4}{l}{$N(\mu; \sigma)$: normal distribution with mean $\mu$ and standard deviation $\sigma$.} \\
\multicolumn{4}{l}{$tN(\mu; \sigma; x_\text{min}; x_\text{max})$: normal distribution with mean $\mu$ and standard deviation $\sigma$, 
truncated from $x_\text{min}$ to $x_\text{max}$.} \\
\end{tabular}
\end{table*}

\begin{table*}[h!]
\centering
\caption{WASP-148 system parameters.
The stellar parameters are obtained from Sect.~\ref{sect_stellar_parameters}.
The transit and orbital parameters are obtained from the Keplerian fit statistics of the joint posterior 
sample (Sect.~\ref{sect_keplerian_characterization}).
In particular, the mean ephemeris
of WASP-148b are obtained from the TTV fit presented in see Sect.~\ref{sect.ttvs}.     
The MAP probability estimates are given together with the standard deviations of 
the marginal distribution following the $\pm$ sign. For some parameters, 
the extremes of the 95\,\% highest density interval are also given in brackets. 
Except for the semiamplitude $K_1$ , which refers to the star, the other orbital parameters 
($\omega$, $\lambda_0$, $T_{\rm 14}$, $b$, $a$...) refer to the planetary orbits.
}            
\begin{minipage}[t]{13.0cm} 
\renewcommand{\footnoterule}{}                          
\begin{tabular}{lccc}        
\hline           
\emph{\hspace{0.3cm}  \vspace{0.1cm} Stellar parameters (Sect.~\ref{sect_stellar_parameters}):} 
 &      &                               \multicolumn{2}{c}{\textbf{WASP-148}}  \\
Stellar mass, $M_*$ &[\Msun] &                                                  \multicolumn{2}{c}{$ 1.00 \pm 0.08 $}  \\
Stellar radius, $R_*$ & [\Rsun] &                                       \multicolumn{2}{c}{$ 1.03 \pm 0.20$}   \\
Stellar density, $\rho_*$ & [g/cm$^3$] &                                        \multicolumn{2}{c}{$ 1.3^{+1.2}_{-0.5}$}   \\
Spectroscopic surface gravity, log\,$g$ & [cgs] &       \multicolumn{2}{c}{$4.40 \pm 0.15$} \\
Effective temperature, $T_{\rm{eff}}$ &[K] &            \multicolumn{2}{c}{5460 $\pm$ 130} \\
Metallicity, $[\rm{Fe/H}]$ & [dex] &                            \multicolumn{2}{c}{+0.11  $\pm$ 0.08} \\
Activity index, $\log{R'_\mathrm{HK}}$ & &              \multicolumn{2}{c}{$-5.09\pm0.11$ } \\
Projected RV, \vsini\ & [km/s]  &              \multicolumn{2}{c}{$3 \pm 2 $ } \\
Rotation period, $P_{\rm rot}$ & [d] & \multicolumn{2}{c}{$26.2\pm1.3$} \\
Barycentric systemic RV,  $\gamma$ & [km/s] &                      \multicolumn{2}{c}{$-5.619 \pm 0.005$} \\
\hline
\emph{\hspace{0.3cm}\vspace{0.1cm} Transit and orbital parameters (Sect.~\ref{sect_keplerian_characterization}):}         & & \textbf{WASP-148b}                  &       \textbf{WASP-148c}                 \\
Orbital period, $P$ & [d] &                             $8.803810 \pm 0.000043$\footnote{Mean ephemeris obtained from the TTV fit presented in see Sect.~\ref{sect.ttvs}.}    &       $34.516 \pm 0.029$      \\
                                &       &                       $[8.803726, 8.803894]$    &   $[34.467, 34.582]$  \\
Argument of periastron, $\omega$ & [deg]        &       $59 \pm 20 $            &       $14 \pm 17$         \\
                        &       &                       $[11.2, 89.4]$  &       $[-20.0, 46.7]$  \\
Time of inferior conjunction or transit, $T_c$ & $[\rm BJD_{\rm UTC}-2\,450\,000$] & $7957.4877 \pm 0.0059^a$  &  $7935.2 \pm 1.1$ \\
                                &       &               $[7957.4761, 7957.4993]$        &    $[7933.2, 7937.9]$   \\      
Time of pericenter passage, $T_p$ & $[\rm BJD_{\rm UTC}-2\;450\,000$] & $7957.00 \pm 0.43$ &  $7931.4 \pm 1.5$ \\
                                &       &               $[7955.84, 7957.52]$    &         $[7928.5, 7934.6]$    \\
Mean longitude at BJD$\,=2\,455\,500$, $\lambda_0$ &  [deg]     &       $27.3 \pm 4.2$ & $214 \pm 17$ \\
                                &       &               $[21.4, 38.4]$  &       $[182.2, 252.2]$ \\
Transit duration, $T_{\rm 14}$ & [hr] &                 $3.016 \pm 0.019$\footnote{Computed under the approximations of Tingley \&\ Sackett~(\cite{tingley05}).}            &       --      \\
                                                &       &               $[2.984, 3.060]$                 &       --      \\
Radius ratio, $R_p/R_{*}$ & &           $0.0807 \pm 0.0007$             &       --        \\
                                &       &               $[0.07943, 0.08224]$    &       --   \\
Normalized semimajor axis, $a/R_*$ &                   &                       $19.8 \pm 1.5$                                &       --       \\
                                &       &               $[17.7, 23.8]$  &       -- \\
Impact parameter, $b$ & &                               $0.046 \pm 0.066$                                &       --      \\
                                        &       &       $[0.000, 0.216]$ & -- \\
Orbital inclination, $i_p$ &  [deg]  &                          $89.80 \pm 0.27$                   &       --      \\
                                        &       &                       $[89.10, 90.00]$         &               --      \\
RV semiamplitude, $K_1$ & [m/s] & $28.7 \pm 2.0$                  &         $25.9 \pm 2.9$          \\
                                        &       &       $[24.5, 32.7]$  &       $[21.4, 33.1]$  \\
$\sqrt{e} \cos{\omega}$ & &                             $0.24 \pm 0.09$          &       $0.58 \pm 0.10$  \\ 
                                        &       &               $[0.033, 0.385]$ &       $[0.344, 0.733]$ \\
$\sqrt{e} \sin{\omega}$ & &                             $0.40 \pm 0.12$          &       $0.14 \pm 0.15$  \\ 
                                        &       &               $[0.05, 0.54]$         &               $[-0.20, 0.38]$ \\
Orbital eccentricity, $e$  &            &       $0.220 \pm 0.063$                       &       $0.359 \pm 0.086$      \\
                                &       &                       $[0.053, 0.306]$ &       $[0.175, 0.517]$  \\
Planet mass (sky-projected), $M_p  \sin i_p$ & $[\Mjup]$  &     $0.291 \pm 0.025$                  &       $0.397 \pm 0.044$  \\
                                &       &       $[0.238, 0.336]$        &       $[0.318, 0.494]$  \\
Planet mass, $M_p$ & $[\Mjup]$  &       $0.291 \pm 0.025$                       &       $0.40 \pm 0.05$  \\
                                &       &       $[0.238, 0.336]$        &       $[0.32, 0.60]$\footnote{see Sect.~\ref{acsct}.}  \\
Planet radius, $R_p$  & $[\Rjup]$  &    $0.722 \pm 0.055$                       &       --       \\
                                &       &       $[0.578, 0.798]$  &   -- \\
Planet density, $\rho_p$ & [g/cm$^3$] &  $0.96 \pm 0.26$  & --  \\
                                &       &       $[0.63, 1.64]$ & -- \\
Orbital semimajor axis, $a$  & [AU] &          $0.0845 \pm 0.0022$& $0.2101 \pm 0.0055$   \\
                        &       &                       $[0.0789, 0.0877]$      &         $[0.1962, 0.2182]$  \\
Planet blackbody equilibrium temperature\footnote{Computed at $a$, assuming a Bond 
albedo of 0.1 and a uniform heat redistribution to the night side.}, 
$T_{\rm eq}$    &       [K] &   $940 \pm 80$    &       $590 \pm 50$ \\
                        &       &                       $[745,1050]$    &         $[470,670]$  \\
$\omega_b - \omega_c$ & [deg] & \multicolumn{2}{c}{$45 \pm 25$} \\
                        &       &       \multicolumn{2}{c}{$[-10.3, 89.9]$} \\
Stellar radius (from WASP-148b transits), $R_*$ & [\Rsun] &                                     \multicolumn{2}{c}{$ 0.918 \pm 0.070$}   \\
                        &       &       \multicolumn{2}{c}{$[0.744, 1.019]$}   \\
Stellar density (from WASP-148b transits), $\rho_{*}$ & [g/cm$^3$] &    \multicolumn{2}{c}{$ 1.89 \pm 0.48$} \\
                        &       &       \multicolumn{2}{c}{$[1.31, 3.16]$} \\
\hline       
\vspace{-0.5cm}
\end{tabular}
\end{minipage}
\label{table.params}  
\end{table*}

\section{Keplerian characterization of the system}
\label{sect_keplerian_characterization}

Here we present a global Keplerian analysis of our photometric and spectroscopic data.
It allows a good fit of the data and reliable  parameters to be derived for the WASP-148 system.
They~are compared  in Sect.~\ref{sect_discussion} with those obtained 
from a Newtonian analysis that takes the mutual interactions of the planets into~account.

\subsection{Models and parameterization}
\label{Sect_Models_and_parametrisation}

The data were fit with the models that are implemented in the {\tt pastis} package (D\'{\i}az et al.~\cite{diaz14}), 
which was originally developed to perform statistical validation of transiting candidates. In brief, the RV time series 
is modeled as a sum of noninteracting Keplerian curves, one per planet. This means that the mutual 
interactions between planets are neglected. To model the transit light curves, {\tt pastis} implements the 
JKTEBOP code (Southworth~\cite{southworth2011}) based on EBOP 
(Nelson \&\ Davis~\cite{nelsondavis1972}, Etzel~\cite{etzel1981}, Popper \&\ 
Etzel~\cite{popperetzel1981}).

The Keplerian curves are parameterized by their period $P$, the semiamplitude $K_1$, 
and two parameters involving the eccentricity and longitude of pericenter, 
$\sqrt{e} \cos(\omega)$ and $\sqrt{e} \sin(\omega)$. For the fifth parameter of 
the curve, we used the time of transit $T_c$ for the inner planet and the mean 
longitude at epoch BJD$\;=2\,455\,500$, $\lambda_0$, for the outer one. 
Finally, we included a systemic RV offset,~$\gamma$. We assumed that the 
RV residuals are independent and normally distributed around zero.
The variance of this distribution is 
equal to the quadratic sum of the inferred uncertainty for each observation 
and an additional term, identical for all data points; this constitutes an 
additional (hyper)parameter of our model, $\sigma^{2}_{RV}$.

The transit light curves were parameterized by the radius ratio 
$k = R_\mathrm{p}/R_\mathrm{s}$, the stellar density, $\rho_*$,
and the impact parameter, $b$. We chose a linear law for the 
stellar limb darkening (LD); this  requires an additional wavelength-dependent 
parameter, the LD coefficient $u_j$, where $j = {1, 2, 3}$ corresponds to the three 
bandpasses used for the observations 
(Johnson-R, og515+kg5, and clear).
The fluxes of each light curve outside of transit 
are additional parameters of the model to allow the normalization to be adjusted.
The RV residuals were assumed to be distributed as the sum of two 
normal distributions, one with a width corresponding to the 
measurement uncertainties, and the other one with a variance 
$\sigma^2_{LC_i}$ different for each light curve $i$.

The data were modeled assuming the presence of two planets only. Initially, we uniquely considered 
the light-curve data,
assuming a constant period for the inner planet and that the outer planet 
did not transit.  However, as reported in Sect.~\ref{sect_FU_photom}, this did not allow us to explain 
the ensemble of the light curves that shows TTVs. We therefore implemented 
a new model in {\tt pastis} that included a time shift for each light curve. This means there is an 
additional model parameter $\delta T_c$ for each light curve. This model accommodated the 
observations and permitted measuring significant departures from 
transit times derived from a constant~period.

We then added the RVs,  modeled as described above, 
and kept the possibility that each transit light curve presented a timing difference to the expected value 
based on a constant ephemeris. This means that the model has an internal inconsistency, as the 
departure from perfect Keplerian motion for the planets is assumed to only be reflected in the 
photometric data. This is justified by the small amplitude of mutual interactions expected for the RVs 
(see Sect.~\ref{sect.ttvs}), as confirmed by our n-body analysis (see~Sect.~\ref{sect_N-body_characterization}).

In these fits, 
the orbital inclination, $i_p$, is not known nor constrained for WASP-148c, 
as well as the longitude of the ascending node, $\Omega$, for both planets (which was set to 0 here).
Some dynamical constraints are placed on these parameters in~Sect.~\ref{acsct},~however.

\subsection{Priors}

The priors chosen for each parameter are presented in Table~\ref{table.priors}. 
These are mostly uninformative priors with some reasonable~bounds. 
Because we did not perform a model comparison, the exact choice of 
bounds is not critical here.
The parameters that use informative priors are the ephemeris parameters for the transiting planet 
for which we employed the knowledge from the transit analysis, the normalizing flux out of transit
for which we chose a normal distribution centered around 1, and the timing offsets used to 
account for possible timing variations, $\delta T_c$. For these parameters, we chose a 
normal distribution centered around zero and with a width of 0.05 day.
This choice is equivalent to adding 
a regularizing term in the likelihood function (Bishop~\cite{bishop2007}) and prevents 
the values the $\delta T_c$ parameters from becoming extremely high by changing 
the value of the period or the nominal transit time accordingly.

\subsection{Posterior sampling}

The posterior distribution was sampled using Markov chain Monte Carlo (MCMC) algorithms. For the first 
analyses we employed the MCMC algorithm that is implemented in {\tt pastis} and described 
by D\'{\i}az et al.~(\cite{diaz14}). This algorithm automatically chooses the parameterization that 
minimizes correlations between the parameters. To do this, the eigenvectors of the empirical covariance 
matrix of the parameters are used to define the directions in which the Markov Chain moves in 
parameter space. For the final run with the full data set, we employed the ensemble sampler described 
by Goodman \&\ Weare~(\cite{goodman10}) that is implemented in the \emcee\ package by Foreman-Mackey et al.~(\cite{foreman13}).

We ran the \emcee\ algorithm with 150 walkers for $2\times10^5$ interactions, and we thinned by a factor of 
100 because of performance (memory) limitations. We removed the first 10\,000 steps, which we assumed to be the 
burn-in period. The walkers exhibited adequate mixing, and the acceptance rate was centered around 0.15. 
All the walkers reside in the same region of parameter space, around a maximum of the posterior density.  
We computed the Geweke~(\cite{geweke1992}) statistics for every parameter and walker and compared the 
first 20\,\% of the chain, with successive fractions of the same size. The results are distributed like a normal 
centered on zero, with a width of around 0.28. All of this indicates that the algorithm likely has converged.

Additionally, we computed the autocorrelation function for the model parameters and selected functions, 
such as the time of inferior conjunction. We did this for each walker individually in an attempt to identify 
chains that presented problems. We did not find any problematic walker, and the mean correlation lengths 
over walkers, defined as the smallest lag for which the autocorrelation is below $1/e$, are below 20 
steps (i.e., 2000 steps from the unthinned chain)~ for all parameters (mean: 13.3, median: 12.3), 
except for $\sqrt{e_b} \sin{\omega_b}$ and the stellar density, $\rho_*$, for which the mean correlation 
lengths are about 23.5 and 20.8, respectively. We therefore have a minimum of 6000 independent 
samples on which to perform the parameter inference.

\subsection{Results}

The samples obtained from the posterior distribution using the MCMC algorithm 
allowed us to derive the maximum-a-posteriori (MAP) estimate for each parameter 
and compute the corresponding model. The MAP model is overplotted on the data 
in Fig.~\ref{fig.lcs}, together with other sample models.
The  derived parameters are reported in Table~\ref{table.params}.

\begin{figure}[b!]
\includegraphics[width=9.2cm, angle=0]{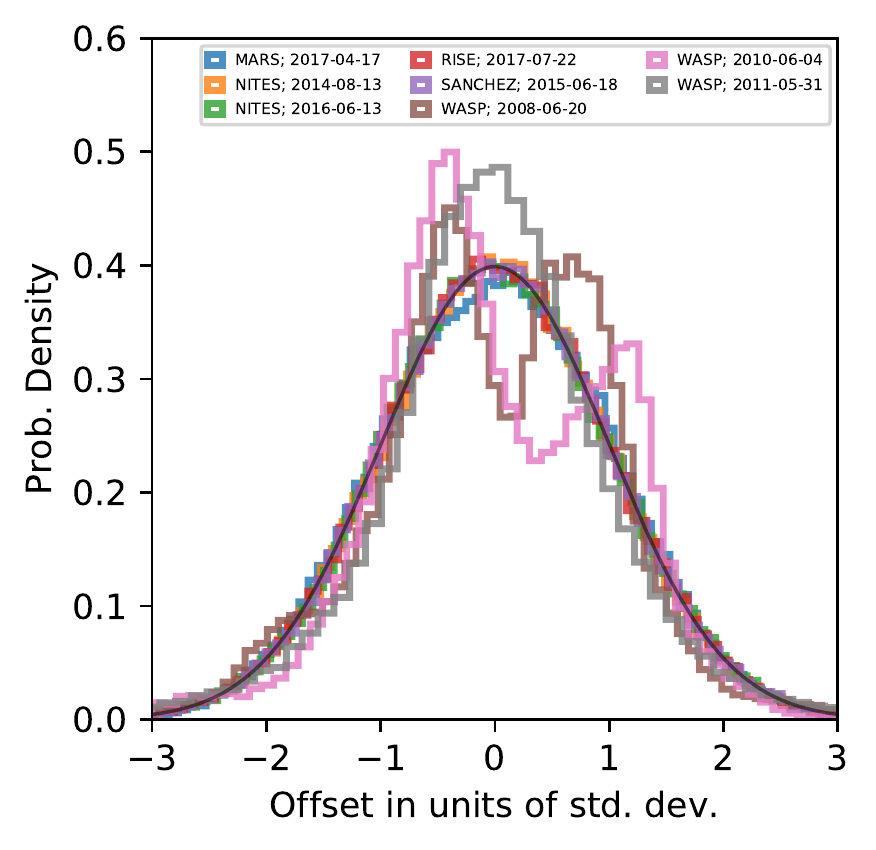}
\caption{Marginal posterior distribution for the WASP-148b transit times for 
each dataset, offset to the mean value of the distribution and normalised using 
the standard deviation. In most cases, the distributions closely resemble the 
standard normal distribution, plotted as a black~curve.}
\label{fig.marginaltcs}
\end{figure}

\subsubsection{Transit-timing variations of WASP-148b}
\label{sect.ttvs}

As reported above, the model we employed to describe the 
data allows for a timing offset for each light curve. 
Based on the posterior samples of the nominal period and 
transit time and on the individual timing offsets, $\delta T_c$, 
we computed the time of inferior conjunction (transit) of 
WASP-148b for each light-curve epoch. Summary statistics 
for the marginal posterior distributions are reported in 
Table~\ref{table.tcs}. Figure~\ref{fig.marginaltcs} shows the 
shape of the marginal posterior distributions. With the exception 
of the transit times of the SuperWASP light curves, which are 
the least precise transit times (Table~\ref{table.tcs}), the inferred marginal 
posterior density functions closely resemble a normal distribution.
This strengthens the reliability of their timings.

With this in mind, we fit a straight line to the transit times, assuming 
normal errors equal to the standard deviations reported above. 
We quadratically added a supplementary error
to the error on each transit time that represents the 
typical amplitude of the possible TTVs, assuming they would present 
sinusoidal variations with time; we show below in Sect.~\ref{sect_fit_ttv} and 
Fig.~\ref{fig_ttv} that this is the case. We adjusted this supplementary 
error in order to have a reduced $\chi^2=1$ to the straight line and found 
that it is equal to 15.1~min.

This leads to a mean ephemeris

\begin{equation}
T^{(n)}_c = 2\,457\,957.4877(59) + n \times 8.803810(43),
\label{equ_ttv} 
\end{equation}
where the number between parentheses represents the error on the 
parameters, and their covariance is 
$\mathrm{cov}\left(T^{(0)}_c, P\right) = 1.76\times10^{-7}$.

\begin{figure}[t!]
\includegraphics{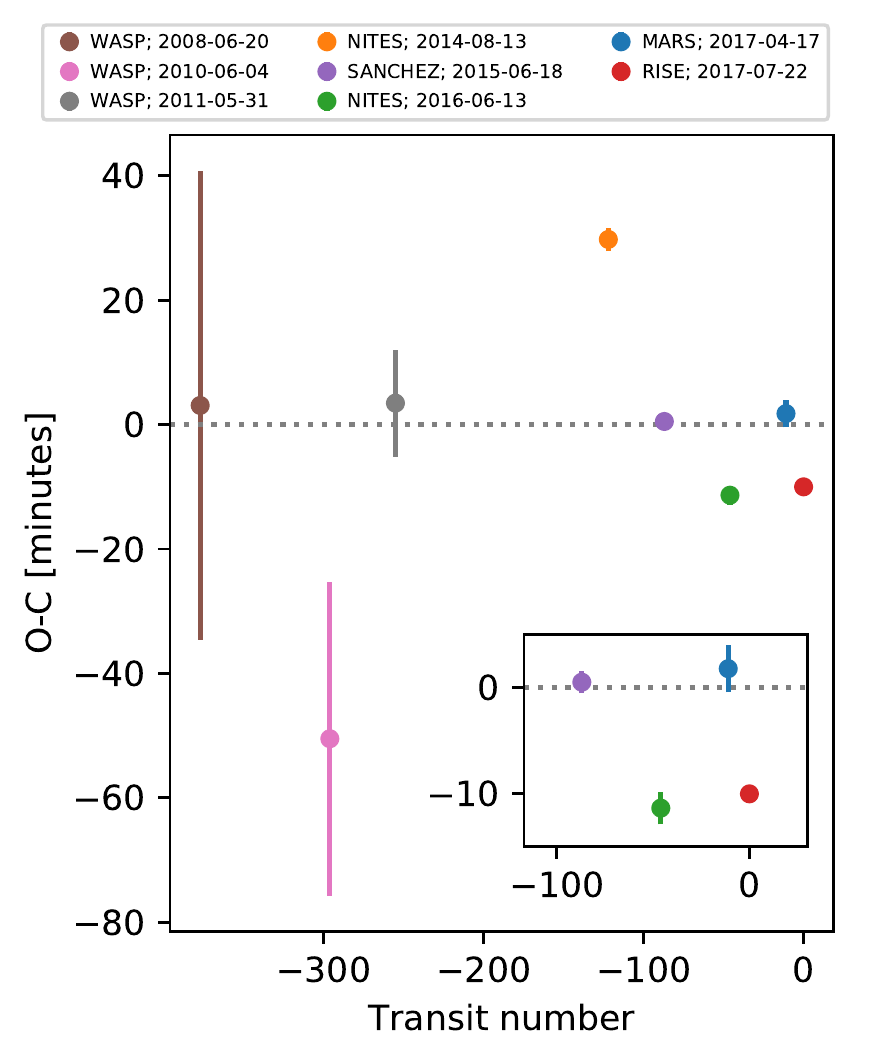}
\caption{Timing residuals of a linear regression model to the inferred WASP-148b  transit times. 
The corresponding constant orbital period is 
   $P=8.803810 \pm 0.000043$~days,
as reported in Table~\ref{table.tcs}.
The inset is a zoom into the region of the last  four transits.
}
\label{fig.oc}
\end{figure}

The residuals of the fit are shown in Fig.~\ref{fig.oc}. 
This confirms that there are 
significant variations with respect to a constant period.
Even when we remove the SuperWASP 
transits (which have large error bars) or the Mars transit (which is partial) from the regression, 
the remaining timing measurements exhibit significant TTVs. 
This means that the TTVs are significantly detected here.

We consider the mean ephemeris as our final ephemeris for WASP-148b. 
It agrees with the ephemeris obtained by the fit above considering Keplerian 
orbits and  a time shift $\delta T_c$ for each light curve.

\begin{figure}[t]
\includegraphics[width=9cm, angle=0]{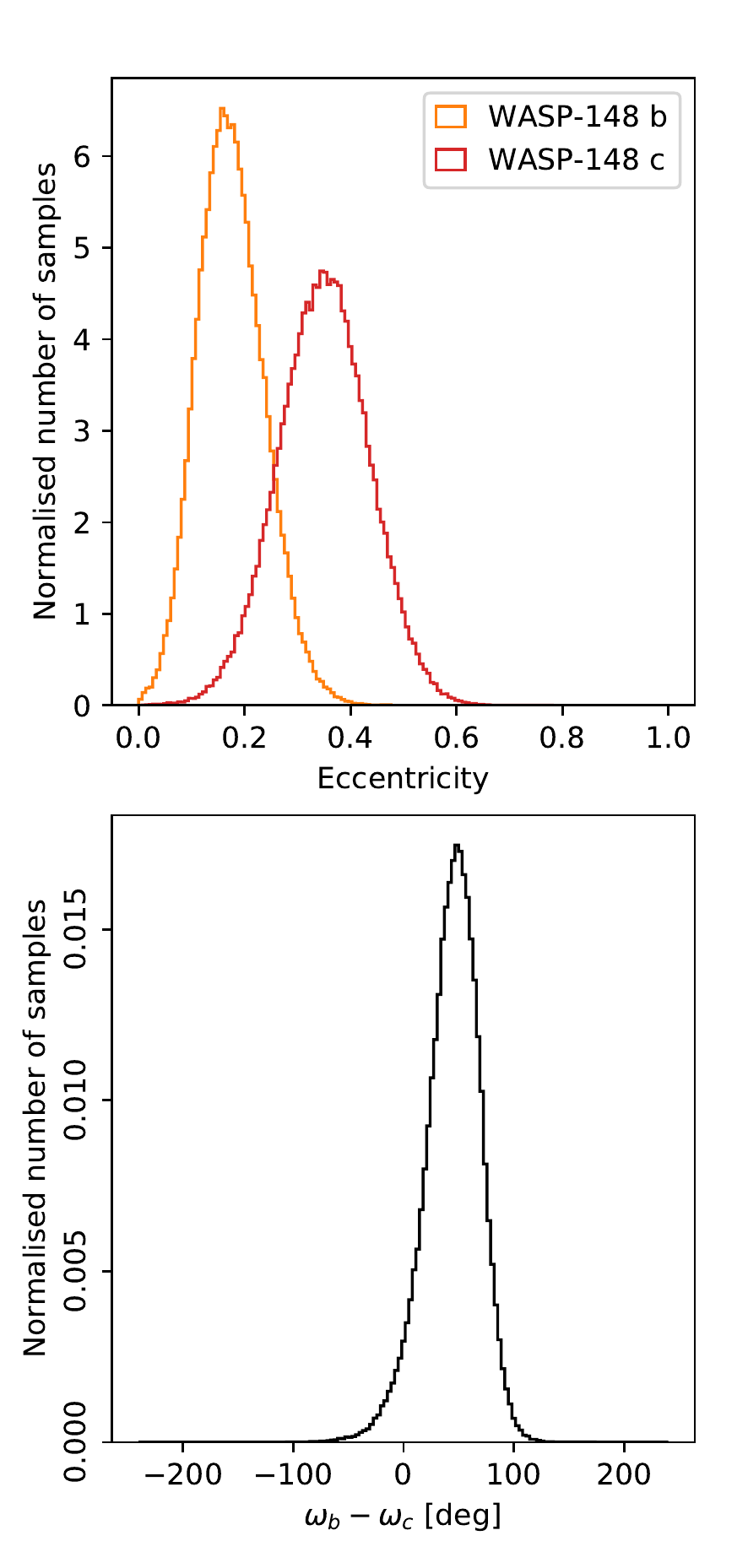}
\caption{Histogram of the marginal posterior samples of the orbital eccentricities of 
WASP-148b and WASP-148c (top) and of the difference in the argument of pericenter (bottom).}
\label{fig.eccdomega}
\end{figure}

\subsubsection{System parameters}
\label{sect_system_parameters}

The results from the sampling of the posterior distribution are summarized in 
Table~\ref{table.params} for some of the model parameters and a number of derived quantities. 
We also provide the 95\,\% highest density interval (HDI), defined as the interval 
containing 95\,\% of the marginal distribution mass, such that all points in the 
interval have probability densities higher than any point outside.
The~reported orbital period and time of transit for WASP-148b are the  averaged periods 
and times obtained from the precedent section. 
The  times for inferior conjunction or transit, $T_c$, and pericenter passage, $T_p$, reported in 
Table~\ref{table.params} are those measured near our most accurate transit (the time for pericenter 
passage is observed with Rise).
For WASP-148c, these parameters are slightly more accurate when they are measured at 
an epoch located in the middle of the SOPHIE observations: 
$T_c = 2\,456\,968.77 \pm 0.90$~BJD and
$T_p = 2\,456\,965.0 \pm 1.3$~BJD.

The results for the nuisance model parameters related to the data sets (flux jitter amplitudes)
are reported in a separate table (Table~\ref{table.tcs}). 
The smallest nuisance model parameter is obtained for the Rise data and is $70\pm 54$~ppm, 
which could be considered as an upper limit for the flux-intrinsic variations of WASP-148.
The derived LD coefficients are $0.93 \pm 0.05$, $0.75 \pm 0.07$, 
and $0.53 \pm 0.05$ for clear, Johnson-R, and og515+kg5, respectively. 
This agrees with expected values in these band passes (e.g., Claret et al.~\cite{claret13}).
The residual dispersion of the RVs is 13.9~m/s, which is slightly higher than the typical 
estimated RV error bars (Sect.~\ref{sect_RV_meas}). The RV jitter we fit to account for this
is $11.1 \pm 1.4$~m/s.

For the derived quantities that required the input from the stellar models, such as the 
mass of the star for the semimajor axis of the orbits, we sampled values of the required 
parameter from normal distributions with a mean and width corresponding to the values reported 
in Sect.~\ref{sect_spectr_charact}.
The transit of WASP-148b allows stellar radius and density to be directly measured as 
$R_*  = 0.92 \pm 0.07$\,\Rsun\ and $\rho_* = 1.89 \pm 0.48$\,g/cm$^3$.
These values are more accurate but agree with those obtained from stellar analysis 
in Sect.~\ref{sect_spectr_charact}  
($R_*  = 1.03 \pm 0.20$\,\Rsun\ and $\rho_* = 1.3^{+1.2}_{-0.5}$\,g/cm$^3$).
Similarly, the $a/R_*$ ratio computed from WASP-148b transits translates into 
$a_b = 0.103 \pm 0.021$\,AU, in agreement with the more accurate value $a_b = 0.0845 \pm 0.0022$
obtained from $M_*$ and the third Kepler law. These agreements reflect the coherence 
of the fit and its~results.

WASP-148b is a hot Jupiter with a mass $M_b = 0.291 \pm 0.025$~M$_\mathrm{Jup}$ and a radius 
$R_b =0.722 \pm 0.055 $\,\Rjup,
translating into a bulk density of  0.95\,g/cm$^3$.
WASP-148c has a sky-projected mass of $M_c  \sin i_p= 0.397 \pm 0.044$~M$_\mathrm{Jup}$.
The orbits of both planets are significantly eccentric, with the MAP estimate of planet b and c being 
$e_b =  0.22 \pm 0.06$  and $e_c = 0.36 \pm 0.09$. 
Based on the determination of the $\omega$ angles, the pericenters might be aligned, 
that is, $\omega_b - \omega_c = 0$ is within the 95\,\%-HDI, but the uncertainties remain too large 
for a significant determination. Figure~\ref{fig.eccdomega} presents the histograms of the 
marginal distributions of the eccentricities and of $\omega_b - \omega_c$.
We also report in  Table~\ref{table.params} the equilibrium temperature $T_{\rm eq}$ of both planets.
It was computed while the planet was at at the semimajor axis of the star, assuming blackbody, 
a Bond albedo of 0.1, and uniform heat redistribution to the planetary nightside.
We find $T_{\rm eq} = 940 \pm 80$\,K and $590 \pm 50$\,K for WASP-148b and WASP-148c, respectively.

The 13.9-m/s dispersion of the RV residuals 
might include the signature of additional planets, such as the possible third planet discussed in 
Sect.~\ref{sect_RVperiodograms}. A three-planet fit does not significantly modify the parameters of 
WASP-148b and WASP-148c.  

Several studies have shown that the Keplerian signature of a single eccentric planet might also be 
fit with a model including two planets on circular (or nearly circular) orbits in 2:1 resonance, 
depending on the RV accuracy and their time sampling  (e.g., Anglada-Escud\'e 
et al.~\cite{anglada10}, Wittenmyer et al.~\cite{wittenmyer13}, K\"urster et al.~\cite{kurster15}).
We therefore attempted to fit our dataset with three-planet models, thus including WASP-148b and 
two outer planets on circular orbits around 17.3 and 34.5~days instead of the eccentric orbit 
of WASP-148c presented above. The quality of this fit is similar to the two-planet fit. 
Except for its eccentricity, the resulting properties of the 34.5 d planet are similar to those of WASP-148c
presented above, in particular its mass. The fit circular 17.3 d planet produces a 
semiamplitude of $9\pm2$\,m/s, which corresponds to a Neptune mass. 
This signal is at the detection limit according to the accuracy of our RV data, 
and the periodograms presented in 
Fig.~\ref{fig_periodograms_RV} show no indication for a signal at about 17 days. 
The planet WASP-148c on a 34.5 d orbit is not placed in doubt, therefore,
but the possibility that its orbit is circular and another planet is present on a 17.3 d circular  
orbit between WASP-148b and WASP-148c  cannot be entirely excluded from the 
available RVs. 
However, we favor here the solution with only two planets, both on eccentric orbits. 
Section~\ref{sect_discussion} below and the TTVs give additional arguments
in favor of eccentric orbits.

\section{Dynamical analyses}
\label{sect_discussion}

The orbital solution given in Table~\ref{table.params} shows a compact system 
($a_b = 0.084 $ and $a_c = 0.210$~AU) with eccentric orbits ($e_b = 0.22$ and $e_c=0.36$).
The stability of the system is not straightforward because the planetary 
masses are on the same order as the mass of Saturn  ($M_b = 0.29$~M$_\mathrm{Jup}$ 
and $M_c  \sin i_p= 0.40$~M$_\mathrm{Jup}$).
In addition, the ratio between the orbital periods, $ P_c / P_b = 3.92 $,  is close to a 4:1 MMR.
As a consequence, mutual gravitational interactions between planets are likely to  be significant.
We study and quantify these dynamical aspects here, with particular focus on the instabilities 
that may arise,  and their effects on the TTVs.

\subsection{N-body characterization of the planetary system}
\label{sect_N-body_characterization}

The solution given in Table~\ref{table.params} was obtained assuming noninteracting 
Keplerian orbits. We first performed an n-body Newtonian fit to the RV data 
taking the mutual interactions of the planets into account. 
We performed this fit using the approach presented by Correia et al.~(\cite{correia10})
and fixing the epoch of WASP-148b transits from Table~\ref{table.params}.

The dispersion of the resulting RV residuals is 14.1~m/s. This is similar to 
the dispersion of the resulting RV residuals of the Keplerian fit 
(13.9~m/s, see Sect.~\ref{sect_system_parameters}), which means that both fits are equivalent 
from this point of view.  The obtained orbital parameters show no significant 
differences with those obtained  for the Keplerian parameters.
This justifies the hypotheses made in Sect.~\ref{sect_keplerian_characterization} 
and the reliability of the parameters reported in Table~\ref{table.params}
over timescales of a few years, which is that of our datasets.
We study the dynamical effects on longer timescales below.

\begin{figure}[b]
\centering
\includegraphics[width=9cm, angle=0]{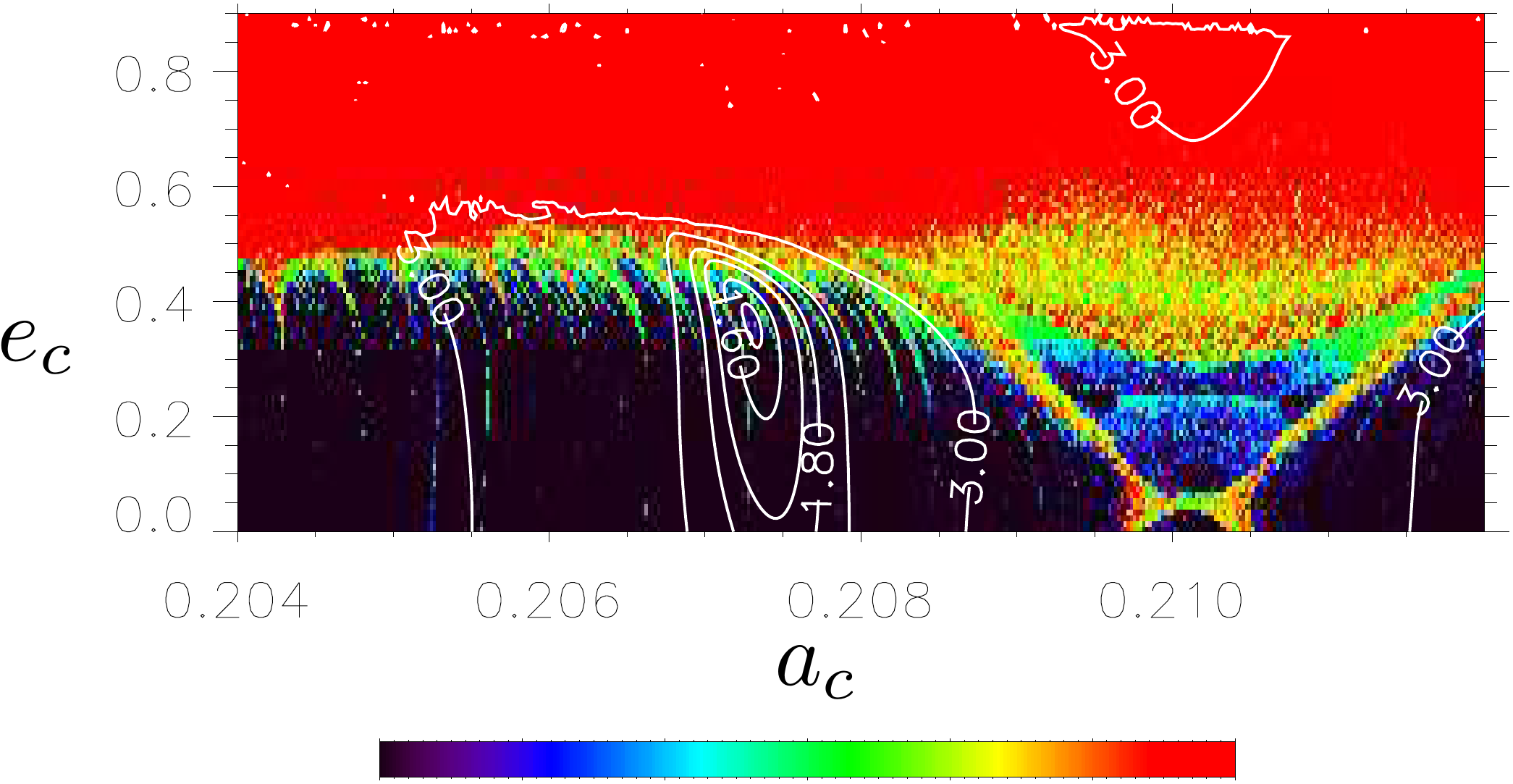}
\caption{Stability analysis of the WASP-148 planetary system, assuming 
coplanar orbits. For fixed initial conditions, the phase space of the system 
is explored by varying the semimajor axis $a_c$ and eccentricity $e_c$ 
of the outer planet WASP-148c.
The step size is $10^{-2}$ in eccentricity and $10^{-3}$ in semimajor axis. 
For each initial conditions, the system is integrated over 50~kyr, and a stability 
criterion is derived with the frequency analysis of the mean longitude.
The chaotic diffusion is measured by the variation in the frequencies. 
The color scale corresponds to values between $-9$  (black) and $-3$ (red) 
for the decimal logarithm of the stability index $D$ used in 
Correia et al.~(\cite{correia10}). 
The red zone corresponds to highly unstable orbits, while  the dark blue 
region can be assumed to be stable on a billion-year timescale.
The reduced-$\chi^2$ level curves of the Newtonian fit are also plotted.}
\label{fig_ae}
\end{figure}

\subsection{Stability analysis}
\label{sect_stability}

We performed a global frequency analysis (Laskar~\cite{laskar90},~\cite{laskar93}) in 
the vicinity of the best fit (Table~\ref{table.params}), in the same way as has been achieved for 
other planetary systems (e.g., Correia et al.~\cite{correia05},~\cite{correia10}).
This allows us to analyse and estimate the stability of the orbital solution.

The system was integrated on a regular 2D mesh of initial conditions, with varying 
semimajor axes and eccentricity of WASP-148c, while the other parameters 
were retained at their nominal values. 
We used the symplectic integrator SABA1064 of Farr\`es et al.~(\cite{farres13}) with a 
step size of $5\times 10^{-3}$~yr and general relativity corrections.
Each initial condition was integrated over 50~kyr, and a stability indicator was 
derived with the frequency analysis of the mean longitude, to be the variation
in the measured mean motion over the two consecutive 25~kyr intervals of time
(for more details, see Couetdic~et al.~\cite{couetdic10}).
For regular motion, there is no significant variation in the mean motion along~the 
trajectory, while it can vary significantly for chaotic~trajectories. 

Figure~\ref{fig_ae}  shows the wide vicinity of the nominal solution, together with the 
reduced-$\chi^2$ level curves (whose minimum gives the best-fit solution 
of the Newtonian fit).
The stability indicator is reported using a color index, where the  
red zones represent the strongly 
chaotic trajectories, and the dark blue zones show the extremely stable trajectories. 
We observe the large 4:1 MMR and its chaotic separatrix.
This system is outside this resonance, 
in a more stable area (dark region).
We hence conclude that the WASP-148 planetary system is stable. 

We also directly tested  the stability of the 
MAP solution from Table~\ref{table.params}
by performing a numerical integration over 1~Gyr. 
As expected, the orbits evolve in a regular way and remain 
stable throughout the simulation.
Nevertheless, the nominal solution is close to an unstable 
region (Fig.~\ref{fig_ae}) because the outer planet is highly eccentric.
This suggests the eccentricity of WASP-148c  might be slightly lower to 
bring the system to an even more stable region.
Overall, this analysis allows further constraints to be placed on the planetary 
parameters by reducing the region of parameter space in which the orbits are stable.

\begin{table}[b]
 \caption{Fundamental frequencies for the nominal orbital solution in
 Table~\ref{table.params}. $n_b$ and $n_c$ are the mean motions, and
 and $g_1$ and $g_2$ are the
 secular frequencies of the pericenters. We assumed coplanar orbits.
 \label{Tdyn1}} 
 \begin{center}
 \begin{tabular}{crrr}
 \hline\hline
      & Frequency   & Period & Angle \\
      & ($^\circ/yr$) & (yr) & (deg) \\
 \hline
 $n_b$ & 14938.848 &             0.0241 &       -5.1139 \\
 $n_c$ &   3807.4413 &           0.0946 &   -125.6927 \\
 $g_1$ &          0.2062 &       1745.6242 & 26.1482 \\
 $g_2$ &          0.6146 & 585.7733 & 168.6632 \\ \hline
\end{tabular}
\end{center}
\end{table}

\begin{figure}[h]
\centering
\includegraphics[width=8.5cm, angle=0]{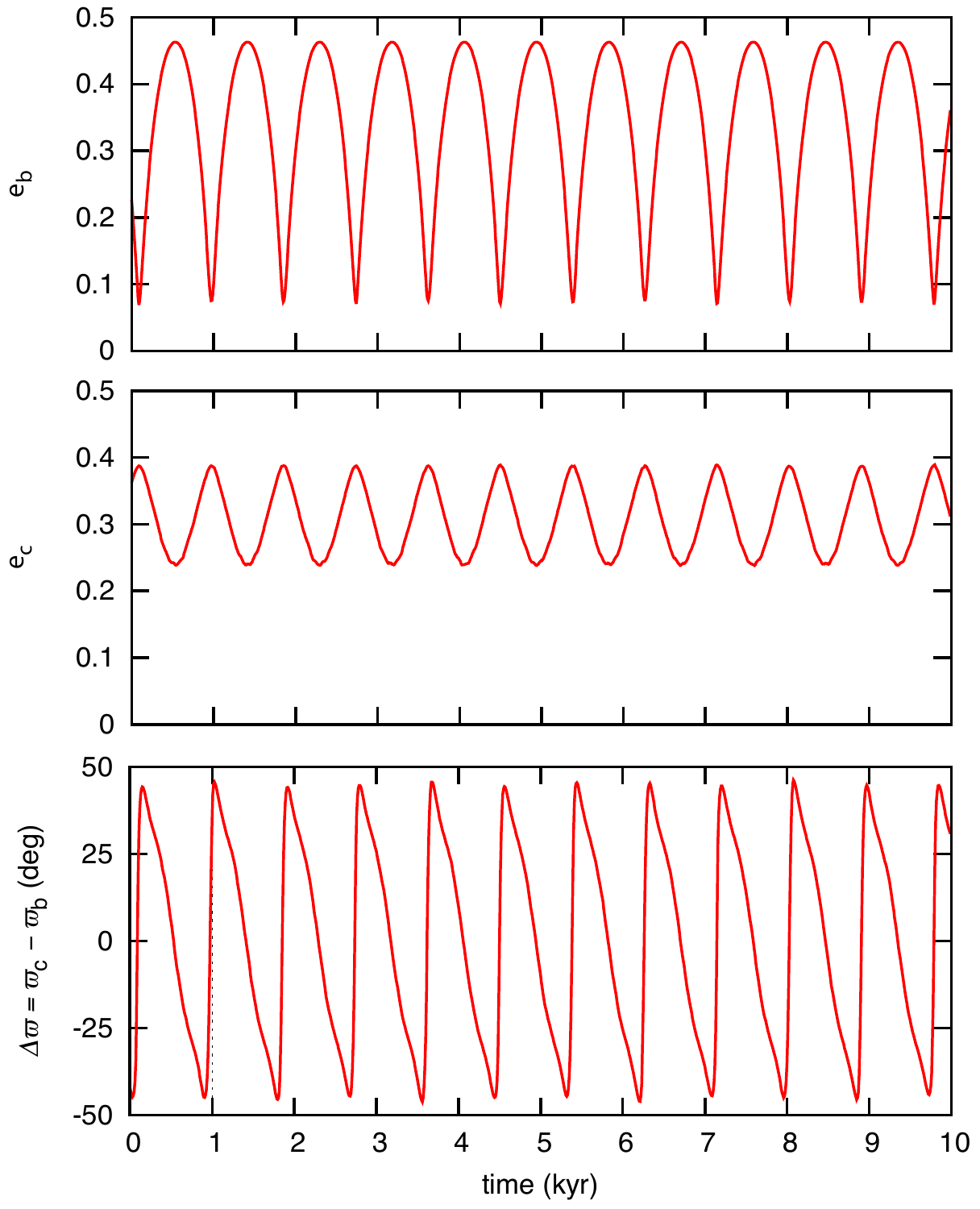}
\caption{Secular evolution of the WASP-148 system, assuming 
coplanar orbits (Table~\ref{table.params}). We show the eccentricity 
of WASP-148b (top) and WASP-148c (middle), and the angle 
$\Delta \omega = \omega_c-\omega_b$ 
(bottom).
These three parameters oscillate with an 881~yr period. 
}
\label{fig_ew}
\end{figure}

In addition, we
performed a frequency analysis of the orbital  solution 
computed over 100~kyr, assuming coplanar orbits.
The fundamental frequencies of the systems are 
the mean motions $n_b$ and $n_c$, and the two secular frequencies of the 
pericenters $g_1$ and $g_2$ (Table\,\ref{Tdyn1}).
Because of the proximity of the two orbits, there is a strong coupling 
within the secular system (see Laskar~\cite{laskar90}).
Planets WASP-148b and c both precess with  the same frequency, $g_1$. 
The two pericenters are thus locked and 
$\Delta \omega = \omega_b - \omega_c$
oscillates around $0^\circ$ (aligned ellipses), with a maximum amplitude 
of about $45^\circ$ (Fig.\,\ref{fig_ew}).
The secular period for the eccentricity and 
$\Delta \omega $ 
oscillations is $2\pi/(g_2-g_1)= 881$~yr. 
On this timescale, the eccentricities of WASP-148b and c therefore roughly oscillate 
between 0.1 and 0.45, and 0.25 and 0.4, respectively. 
Whereas most hot Jupiters have circularized orbits as a result of tidal dissipation  
(e.g., Dawson \&\ Johnson~\cite{dawson18}), the eccentricity of
WASP-148b might at least partially be explained by its interactions with WASP-148c.

\subsection{Additional constraints}
\label{acsct}

Because WASP-148b transits its host star, we are able to determine its inclination to the 
line of sight $i_b = 89.80^\circ \pm 0.27^\circ$ (Table~\ref{table.params}).
As a result, the system is left with only two undetermined parameters: the orbital 
inclination of the outer planet, $i_c$, and the difference between the longitude of the 
ascending nodes, $\Delta \Omega  = \Omega_c - \Omega_b$. 
The longitude of the ascending node of WASP-148b can be fixed at any value, therefore for 
simplicity, we fixed $\Omega_b=0^\circ$ and $\Delta \Omega = \Omega_c$. We thus 
built a 2D stability map for the two unknown parameters to determine how dynamics can 
constrain their possible values.

\begin{figure}[t]
\centering
\includegraphics[width=9cm, angle=0]{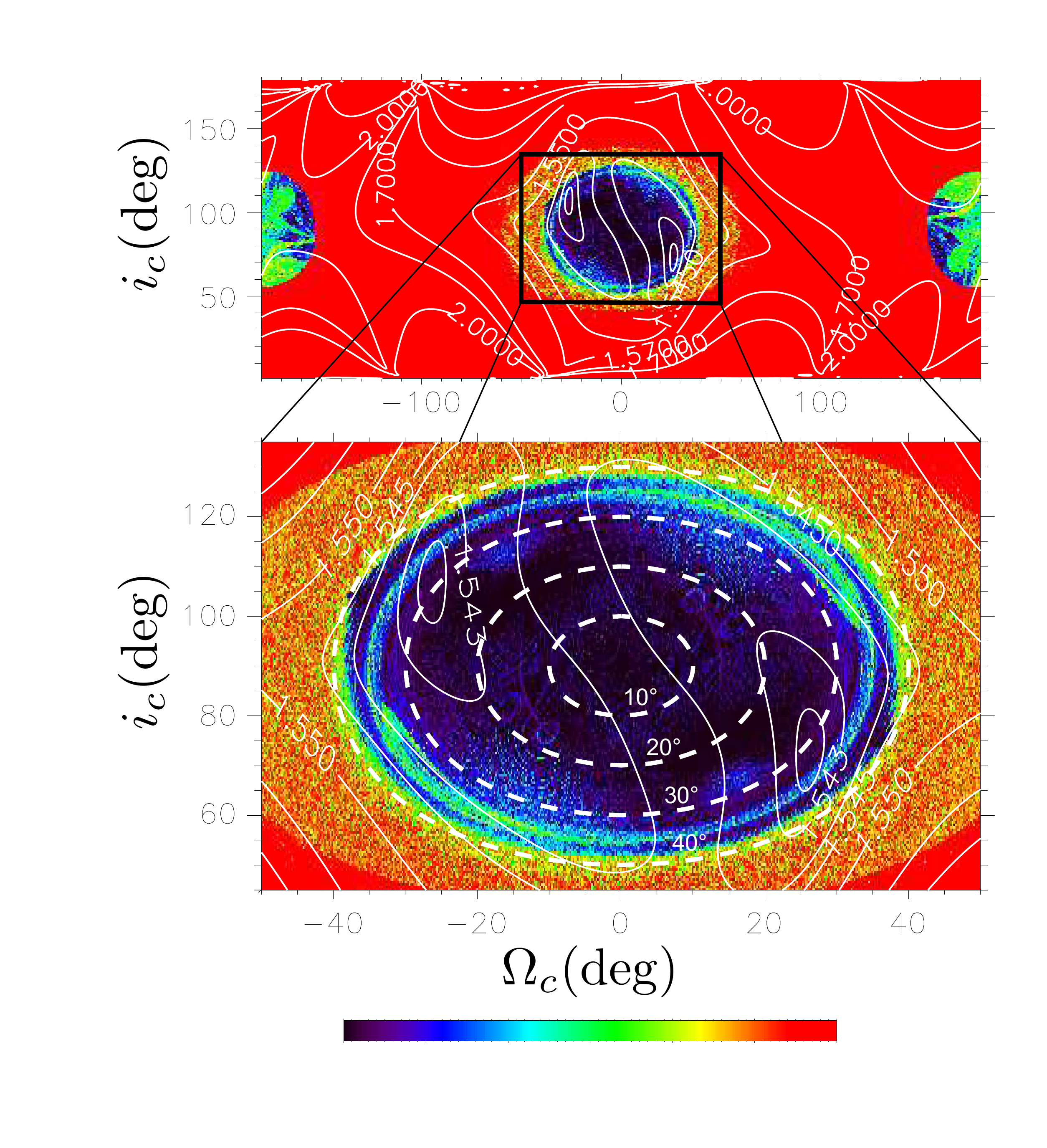}
\caption{Global stability analysis of the WASP-148 planetary system. 
We fixed all orbital parameters of the solution shown in Table~\ref{table.params} 
and varied the only two unconstrained parameters, both from the outer planet WASP148c: 
the longitude of its ascending node, $\Omega_c$, and its inclination, $i_c$. 
The upper panel shows the whole $(i_c, \Omega_c)$ domain, whereas the lower panel 
zooms ino the most stable regions. The step size was $0.2^\circ$ in the node and $0.5^\circ$ in the 
inclination. For each initial condition, the system was integrated over 50~kyr and a 
stability criterion was derived with the frequency analysis of the mean longitude. 
Because of the dynamical invariant, the  figure is symmetric with respect to the 
$(i_c=90^\circ, \Omega_c=0)$ center. 
White dashed curves give the isolines of constant mutual inclination $I=10^\circ, 20^\circ, 
30^\circ, \text{and } 40^\circ$. The color scale corresponds to values between $-9$  (black) and $-3$ (red) 
for the decimal logarithm of the stability index $D$ used in 
Correia et al.~(\cite{correia10}). 
The red zones correspond to highly unstable orbits, while  the dark blue 
region can be assumed to be stable on a billion-year timescale.
The reduced-$\chi^2$ level curves of the Newtonian fit are also plotted in white.
}
\label{fig_i2o2}
\end{figure}

Figure~\ref{fig_i2o2} explores the  stability in the $(i_c, \Omega_c)$ domain by keeping the 
remaining parameters fixed at the values shown in Table~\ref{table.params}.
We also show the reduced-$\chi^2$ level curves. 
They present a minimum for $i_c=73^\circ, \Omega_c = 26^\circ$
(and by symmetry $i_c=107^\circ, \Omega_c = -26^\circ$),
but the contour levels do not place strong constraints on the determination of this minimum
with the available data. We are therefore unable to determine these 
parameters from our Newtonian fit at present (see, e.g., Correia et al.~\cite{correia10}).
However, there is only one subset of $(i_c, \Omega_c)$ values for which 
the system can be stable: 
$55^\circ \lesssim i_c \le 125^\circ$ and $| \Omega_c | \lesssim 35^\circ$
(Fig.~\ref{fig_i2o2}, lower panel).
We note that a stable zone for retrograde orbits also exists (around 
$\Omega_c = 180^\circ$;  
Fig.~\ref{fig_i2o2}, upper panel), but it is excluded because the RVs indicate a prograde orbit 
of WASP-148c by comparison to WASP-148b.
These dynamical constraints have direct consequences on the determination of 
the true mass of WASP-148c, $M_c$,  despite the absence of transit detection. 
Using the 95\,\%-HDI upper limit 
$M_c \sin i_c < 0.49$~M$_\mathrm{Jup}$ (Table~\ref{table.params}), 
we conclude that $M_c \le 0.60$~M$_\mathrm{Jup}$.  
This is a stringent upper limit on the true mass of an exoplanet detected from RV~alone.

Another constraint can be derived for the mutual inclination between orbital planes, $I$.
Assuming $i_b = 90^\circ$, we have
\begin{equation}
\cos{I} = \sin{i_c} \cos{\Omega_c} \ . \label{eq4} 
\end{equation}
In Fig.~\ref{fig_i2o2} we plot the lines of constant mutual inclination, which describe circles 
around $(i_c=0^\circ, \Omega_c=0^\circ)$. We conclude that all stable areas correspond to
$I \lesssim 35^\circ$, so that the orbital plane of the two planets cannot have a mutual inclination 
higher than this~value.

\subsection{Transit-timing variations}
\label{sect_fit_ttv}

As shown in Sects.~\ref{sect_observations} and~\ref{sect.ttvs}, 
whereas transits of WASP-148c have not been detected nor ruled out
with the existing data,  WASP-148b does transit 
in front of the host star and shows significant TTVs.
These TTVs are likely to be   produced by gravitational interactions with WASP-148c
and can be used to constrain its orbit, in particular, the eccentricity  or the 
true mass (e.g., Lissauer at al.~\cite{lissauer11}).
We model here those TTVs.

\begin{figure}[b]
\centering
\includegraphics[width=9cm, angle=0]{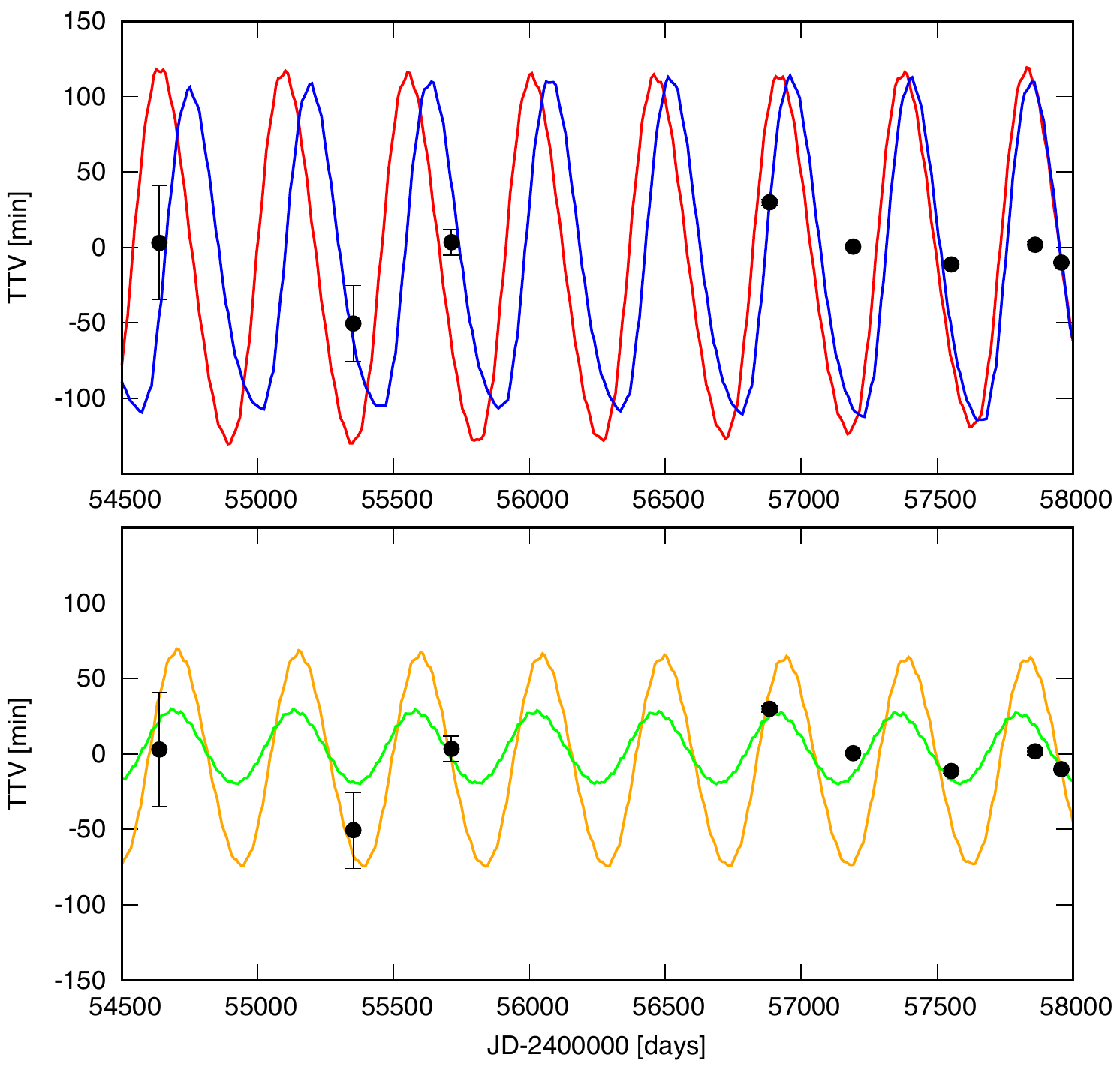}
\caption{Transit-timing variations for the WASP-148b planet. In the upper panel,
the curves correspond to numerical 
simulations for the solution shown in Table~\ref{table.params}
with $i_c = 90^\circ, \Omega_c = 0^\circ$ (coplanar orbits, in red), 
or with $i_c = 60^\circ, \Omega_c = 0^\circ$  
(mutual inclination $I=30^\circ$, in blue). The dots correspond to the 
observed TTVs (Table~\ref{table.tcs}). For comparison, 
the lower panel presents curves corresponding to coplanar solutions with lower eccentricities, 
within our 95\,\%-HDI:
 $e_b = 0.18$, $e_c = 0.29$ (in orange) or  $e_b = 0.11$, $e_c = 0.21$ (in~green).
}
\label{fig_ttv}
\end{figure}

Using the MAP solution from Table~\ref{table.params} and fixing 
$\Omega_c=0^\circ$, we generated the TTVs 
for two different configurations: one with $i_c = 90^\circ$, corresponding to a coplanar system, 
and another with $i_c = 60^\circ$, corresponding to a mutual inclination $I=30^\circ$. 
In the upper panel of 
Fig.~\ref{fig_ttv} we show the variations corresponding to each solution.
Over our ten-year observations, 
they present a sinusoidal shape with a period of about 460~days and an 
amplitude of about two hours. This is the expected shape in a configuration of 
two planets like this near MMR, and it agrees with the assumption made in Sect.~\ref{sect.ttvs}.

Figure~\ref{fig_ttv} also shows the  measured TTVs  as reported in Table~\ref{table.tcs}.
The two-hour amplitude of the computed TTVs is somewhat  larger than 
the observations. 
The modest agreement might partially be explained as follows: we did not fit the TTVs 
simultaneously with  the other data, but instead measured the TTVs and then fit them. 
It might also be partially due to the uncertainty in some of the orbital parameters,
in particular, the eccentricities and the arguments of the pericenter.
The lower panel of 
Fig.~\ref{fig_ttv}  presents the predicted TTVs assuming lower eccentricities, 
within the 95\,\% HDI.
Here their amplitudes agree better with the observations, 
in particular, the model assuming eccentricities of 0.11  and 0.21
for WASP-148b and c, respectively.
The precision and the number of photometric measurements currently available for the 
WASP-148 system, together with the fact that TESS will soon observe it (Sect.~\ref{sect_conclusion}),
do not justify running an exhaustive search for a best-fit solution, 
and hence reduce the uncertainty in the parameters ($e,\omega)$ and constrain the 
unknown parameters~$(i_c,\Omega_c)$.

Finally, we note that when we assume that both orbits are circular, the amplitude of the WASP-148b TTVs 
would be negligible, on the order of a few seconds or below. The detection of TTVs therefore supports 
the assumption that the orbits are eccentric (Sect.~\ref{sect_system_parameters}).
If WASP-148c were eventually be discovered to be transiting its host star, it should also present TTVs.
They would be anticorrelated with those of WASP-148b, but have a larger 
amplitude because its orbital period is longer, which has also been observed for Kepler-9, for example.

\section{Conclusions}
\label{sect_conclusion}

We have presented the discovery and characterization of the WASP-148 
exoplanetary system. This is based on ten years of photometric and spectroscopic 
observations and their Keplerian and Newtonian analyses. 
The system includes a 0.29-\Mjup, 
0.72-\Rjup\  hot Jupiter transiting its star every 8.80~days, 
and an outer planet  that is apparently not transiting with a period of 34.5~days
and a sky-projected mass of 0.40-\Mjup\  (true mass below 0.60~M$_\mathrm{Jup}$).
The planetary equilibrium temperatures are 940\,K and 590\,K, respectively.
The orbits of both planets are eccentric and have 
a mutual inclination below $35^\circ$. They present significant gravitational 
interactions due to their period ratio near the 4:1 MMR. 
This orbital configuration is stable, but shows significant deviations from purely Keplerian orbits.
In particular, the inner planet exhibits TTVs of about 15~minutes.

This configuration makes WASP-148 a unique case. 
As systems with the greatest similarities with WASP-148, but still with significant differences, 
we can cite Kepler-9, Kepler-277, or TOI-216.
Kepler-9 (Holman et al.~\cite{holman10}) is briefly described above in Sect.~\ref{sect_intro}.
Kepler-277 hosts two transiting planets with radii that are about three times the Earth radius and orbital 
periods of 17.3 and 33.0~days (Wu \&\ Lithwick~\cite{wu13}, Xie~\cite{xie14}). 
This configuration causes TTVs of a few 
minutes, which implies masses similar to those of the WASP-148 planets. The Kepler-277 planets are thus 
particularly dense, but their masses remain poorly known as the eccentricities of the orbits are 
not measured.
TOI-216 hosts two transiting giant planets with periods of 17.1 and 34.6~days
(Kipping et al.~\cite{kipping19}, Dawson et al.~\cite{dawson19}). TTVs of a few 
minutes are detected and allow masses of 0.56 and 0.08~\Mjup\ to be evaluated, respectively.
Here the eccentricities and the planetary densities are low.
These three systems show TTVs and are near the 2:1 MMR. Their planets are located 
in the period valley, which is a domain of orbital periods between 10 and 100 days that is known to 
be sparse in giant planets (e.g., Udry et al.~\cite{udry03}, Santerne et al~\cite{santerne16}).
Their MMR configurations might be linked to the fact they are present in the valley. 

WASP-148 is in a similar configuration, but here close to the 4:1 MMR, and WASP-148b 
shares properties with standard hot Jupiters. A significant difference between WASP-148 
and the three systems above is the eccentricity of its planets, which here are significantly 
different from zero.
This is also a significant difference of WASP-148b to other hot Jupiters, most of whose 
orbits are circular.

The NASA Transiting Exoplanet Survey Satellite (TESS) secures an all-sky photometric survey 
to detect planetary transits in front of bright stars (Ricker et al.~\cite{ricker15}).
Its camera 2 will observe WASP-148 in its sectors 24, 25, and 26
from 2020 April 16 to  July 4. It will thus observe nine consecutive transits of WASP-148b, which means that it will 
significantly increase the number of available transits and dramatically improve their accuracy. 
This will improve the transit parameters, and more importantly, it is expected to allow the TTVs
to be confirmed and refined.
Together with the fact that they will be obtained consecutively and with a single instrument, 
the higher quality of these new transit light curves 
will allow the dynamical model of the WASP-148 system to be refined, 
and in particular, a full TTV analysis can be achieved. 
In addition, TESS will cover two inferior conjunctions of WASP-148c 
(on 2020 May 1  and June 4).            
This might reveal that this outer planet does transit its host star (and also presents TTVs), 
or it might not.  If WASP-148b and c are coplanar, there is a 99.4\%\ probability for WASP-148c to transit.
If the mutual inclination between both orbits 
is $I=1^\circ$ or $5^\circ$, this probability falls to 63\,\%\ and 4\,\%, respectively.

The future TESS 79-day continuous observation may also reveal additional 
planets in the system at short or long periods, in particular, small-size planets that 
so far could not be significantly detected in our RV data but may be found in an accurate 
space-based light curve. We will also continue our RV follow-up of the system in order 
to refine the system parameters, in particular, the planetary eccentricities, and to possibly 
confirm and characterize the possible long-period planet 
that shows a detection hint in our current RV dataset. Spectroscopic 
observations of a transit of WASP-148b may also be secured to measure its obliquity
(e.g., Winn et al.~\cite{winn09}, H\'ebrard et al.~\cite{hebrard11}).
The amplitude of the Rossiter-McLaughlin anomaly might be about 10\,m/s or larger,
which means that it can be reached by several spectrographs.

\begin{acknowledgements}
We thank the Observatoire de Haute-Provence (CNRS) staff for its support.
This work was  supported by the ``Programme National de Plan\'etologie'' (PNP) of CNRS/INSU, and
the CNRS-PICS program between France and Argentina  (PICS07826).
A.C.M.C. acknowledges support by 
CFisUC strategic project (UID/FIS/04564/2019),
ENGAGE SKA (POCI-01-0145-FEDER-022217), and
PHOBOS (POCI-01-0145-FEDER-029932),
funded by COMPETE 2020 and FCT, Portugal.
O.D.S.D. acknowledges support from FCT and FEDER/COMPETE2020 in the form of the 
work contract DL 57/2016/CP1364/CT0004 and the projects EPIC (PTDC/FIS-AST/28953/2017
\&\ POCI-01-0145-FEDER-028953) and GEANES (PTDC/FIS-AST/32113/2017 - 
POCI-01-0145-FEDER-032113).
This work was granted access to the HPC resources of MesoPSL financed
by the R\'egion \^Ile-de-France and the project Equip@Meso 
of the programme Investissements d'Avenir supervised
by the Agence Nationale pour la Recherche (ANR-10-EQPX-29-01).
This work is partly financed by the Spanish Ministry of Economics and 
Competitiveness through project ESP2016-80435-C2-2-R and 
PGC2018-098153-B-C31.
This article is based on observations made with different facilities, including
the SOPHIE spectrograph at the 1.93-m telescope of Observatoire Haute-Provence, France, 
the SuperWASP-North instrument located on La Palma in the Canary Islands, Spain, 
the Near Infra-red Transiting ExoplanetS (NITES) Telescope located at La Palma in the 
Canary Islands, Spain, 
the FastCam and Wide-FastCam instruments, at Telescopio Carlos S{\'a}nchez operated 
on the island of Tenerife by the IAC in the Spanish Observatorio del Teide, 
and the Observatoire \textit{Hubert Reeves} in Mars, France.
This paper was finalized during the COVID-19 pandemic.
\end{acknowledgements}


\begin{thebibliography}{}

\bibitem[2005]{agol05} 
Agol, E., Steffen, ., Sari, R., et al. 2005, \mnras, 359, 567

\bibitem[2009]{almenara09} 
Almenara, J. M., Deeg, H. J., Aigrain, S., et al. 2009, \aap, 506, 337

\bibitem[2010]{anglada10} 
Anglada-Escud\'e, G., L\'opez-Morales, M., Chambers, J. E. 2010, \apj, 709, 168

\bibitem[2009]{asplund09} 
Asplund, Ma., Grevesse, N., Sauval, A. J., et al. 2009, ARA\&A, 47, 481

\bibitem[2020]{baluev20} 
Baluev, R., Sokov, E., Hoyer, S., et al. 2020, \mnras, 496, L11

\bibitem[1996]{baranne96} 
Baranne, A., Queloz, D., Mayor, M., et al. 1994, \aaps, 119, 373

\bibitem[2014]{barros14} 
Barros, S. C. C., D\'{\i}az, R. F., Santerne, A., et al.2014, \aap, 561, L1

\bibitem[2015]{becker15} 
Becker, J. C., Vanderburg, A., Adams, F. C., et al. 2015, \apj, 812, L18

\bibitem[2017]{bianchi17} 
Bianchi, L.,  Shiao, B.,  Thilker, D. 2017,  \apjs, 230, 24

\bibitem[2014]{birkby14} 
Birkby, J. L., Cappetta, M., Cruz, P., et al. 2014, \mnras, 440, 1470

\bibitem[2007]{bishop2007} 
Bishop, C. M. 2007, Pattern Recognition and Machine Learning (Information Science and Statistics), 
1st edn. (Springer)

\bibitem[2009]{boisse09} 
Boisse, I., Moutou, C., Vidal-Madjar, A., et al. 2009, \aap, 495, 959

\bibitem[2010]{boisse10} 
Boisse, I., Eggenberger, A., Santos, N. C., et al. 2010, \aap, 523, A88

\bibitem[2009a]{bouchy09a} 
Bouchy, F., H\'ebrard, G., Udry, S., et al. 2009a, \aap, 505, 853

\bibitem[2009b]{bouchy09b} 
Bouchy, F., Isambert, J., Lovis, C., et al. 2009b, EAS Pub. Series, 37, 247

\bibitem[2013]{bouchy13} 
Bouchy, F., D\'{\i}az, R. F., H\'ebrard, G., et al. 2013, \aap, 549, A49

\bibitem[2019]{bouma19} 
Bouma, L. G., Winn, J. N., Baxter, C, et al.  2019, \aj, 157, 217

\bibitem[2020]{bouma20} 
Bouma, L. G., Winn, J. N., Howard, A. W., et al. 2020, \apj, 893, L29

\bibitem[2000]{charbonneau00} 
Charbonneau, D., Brown, T. M., Latham, D. W., et al.  2000, \apj, 529, L45

\bibitem[2017]{christiansen17} 
Christiansen, J. L., Vanderburg, A., Burt, J., et al. 2017, /aj, 154, 122

\bibitem[2013]{claret13} 
Claret, A., Hauschildt, P. H., Witte, S. 2013, \aap, 552, A16 

\bibitem[2006]{cameron06} 
Collier Cameron, A., Pollacco, D., Street, R. A., et al. 2006, \mnras, 373, 799

\bibitem[2020]{cooke20} 
Cooke, B. F., Pollacco, D., Almleaku, Y., et al. 2020, \aj, 159, 255

\bibitem[2005]{correia05} 
Correia, A. C. M., Udry, S., Mayor, M., et al. 2005, \aap, 440, 751

\bibitem[2010]{correia10} 
Correia, A. C. M., Couetdic, J., Laskar, J., et al. 2010, \aap, 511, A21

\bibitem[2010]{couetdic10} 
Couetdic, J., Laskar, J., Correia, A. C. M., et al. 
2010, \aap, 519, A10

\bibitem[2003]{cutri03} 
Cutri, R. M.,  Skrutskie, M. F.,  van Dyk, S., et al. 2003, yCat 2246

\bibitem[2018]{dawson18} 
Dawson, R. I., Johnson, J. A. 2018, ARA\&A, 56, 175

\bibitem[2019]{dawson19} 
Dawson, R. I., Huang, C. X.. Lissauer, J. J., et al. 2019, \aj, 158, 65

\bibitem[2018]{demangeon18}
Demangeon, O. D. S., Faedi, F., H\'ebrard, G., et al. 2018, \aap, 610, A63

\bibitem[2008]{diaz08} 
D{\'{\i}}az, R.~F., Rojo, P., Melita, M., et~al. 2014, \apj, 682, L49

\bibitem[2014]{diaz14} 
D{\'{\i}}az, R.~F., Almenara, J.~M., Santerne, A., et~al. 2014, \mnras, 441, 983

\bibitem[2013]{doyle13} 
Doyle, A. P., Smalley, B., Maxted, P. F. L., et al., \mnras, 428, 3164

\bibitem[2014]{doyle14} 
Doyle, A. P., Davies, G. R., Smalley, B., et al. 
2014, \mnras, 444, 3592

\bibitem[1981]{etzel1981} 
Etzel, P. B. 1981, in Photometric and Spectroscopic Binary Systems, ed. E. B.
Carling \&\ Z. Kopal, 111

\bibitem[2013]{farres13} 
Farr\`es, A., Laskar, J., Blanes, S., et al. 2013, 
Celestial Mechanics and Dynamical Astronomy, 116, 141

\bibitem[2013]{foreman13} 
Foreman-Mackey, D., Hogg, D.~W., Lang, D., et al.  
2013, \pasp, 125, 306
  
\bibitem[2013]{fossati13} 
Fossati, L., Ayres, T. R., Haswell, C. A., et al. 2013, \apj, 766, L20

\bibitem[2016]{gaia16} 
Gaia Collaboration, 2016, \aap, 595, A1

\bibitem[2018]{gaia18} 
Gaia Collaboration, 2018, \aap, 616, A1

\bibitem[1992]{geweke1992} 
Geweke, J. 1992, in Bayesian Statistics 4, ed. A.D.J.M.~Bernardo, J.O.~Berger
  \& A.~Smith (Oxford University Press)

\bibitem[2017]{gillon17}
Gillon, M., Triaud, A. H. M. J., Demory, B.-O., et al.  2017, Nature, 542, 456

\bibitem[2010]{goodman10} 
Goodman, J. \& Weare, J. 2010, 
Communications in applied mathematics and computational science, 5, 65

\bibitem[2012]{haswell12} 
Haswell, C. A., Fossati, L., Ayres, T., et al. 2012, \apj, 760, 79

\bibitem[2019]{haswell19} 
Haswell, C. A., Staab, D., Barnes, J. R., et al. 2019, NatAs, 4, 408

\bibitem[2016]{hay16} 
Hay, K. L., Collier-Cameron, A., Doyle, A. P., et al. 2016, \mnras, 463, 3276

\bibitem[2008]{hebrard08} 
H\'ebrard, G., Bouchy, F., Pont, F., et al. 2008, \aap, 481, 52

\bibitem[2011]{hebrard11} 
H\'ebrard, G., Ehrenreich, D., Bouchy, F., et al. 2011, \aap, 527, L11

\bibitem[2013]{hebrard13} 
H\'ebrard, G., Collier Cameron, A., Brown, D. J. A., et al. 2013, \aap, 549, A134

\bibitem[2014]{hebrard14} 
H\'ebrard, G., Santerne, A., Montagnier, G., et al. 2014, \aap, 572, A93

\bibitem[2015]{henden15} 
Henden, A. A.,  Levine, S.  Terrell, D.  Welch, D. L. 2015,  BAAS, 225, 336.16

\bibitem[2000]{hog00} 
H\o g, E.,  Fabricius, C.,  Makarov, V. V., et al. 2000, \aap 355, L27

\bibitem[2005]{holman05} 
Holman, M. J., Murray, N. W. 2005, Science, 307, 1288

\bibitem[2010]{holman10} 
Holman, M. J., Fabrycky, D. C., Ragozzine, D., et al. 2010, Science, 330, 51

\bibitem[2019]{kiefer19} 
Kiefer, F., H\'ebrard, G., Sahlmann, J., et al. 2019, \aap, 631, A125

\bibitem[2019]{kipping19} 
Kipping, D., Nesvorn\'y, D., Hartman, J., et al. 2019, \mnras, 486, 4980

\bibitem[2015]{kurster15} 
K\"urster, M., Trifonov, T., Reffert, S., et al. 2015, \aap, 577, A.103

\bibitem[1993]{kurucz93} 
Kurucz, R. 1993, Cambridge, Mass.: Smithsonian Astrophysical Observatory, 13

\bibitem[1990]{laskar90} 
Laskar, J. 1990, Icarus, 88, 266

\bibitem[1993]{laskar93} 
Laskar, J. 1993, Physica D Nonlinear Phenomena, 67, 257

\bibitem[2011]{lissauer11} 
Lissauer, J. J., Fabrycky, D. C., Ford, E. B., et al. 2011, Nature, 470, 53

\bibitem[2010]{maciejewski10} 
Maciejewski, G., Dimitrov, D., Neuh\"auser, R., et al. 2010, \mnras, 407, 2625

\bibitem[2016]{maciejewski16} 
Maciejewski, G., Dimitrov, D., Fern\'andez, M., et al., 2016, \aap, 588, L6

\bibitem[2014]{mccormac14} 
McCormac, J., Skillen, I., Pollacco, D. et al. 2014, \mnras, 438, 3383

\bibitem[2011]{maxted11} 
Maxted P. F. L., Anderson D. R., Collier Cameron A., et al. 2011, \pasp, 123,~547

\bibitem[2015]{motalebi15} 
Motalebi, F., Udry, S., Gillon, M., et al. 2015, \aap, 584, A72

\bibitem[1972]{nelsondavis1972} 
Nelson, B., Davis, W. D. 1972, \apj, 174, 617

\bibitem[2012]{nesvorny12} 
Nesvorn\'y, D.,  Kipping, D., Buchhave, L. A., et al.  2012, Science, 336, 1133

\bibitem[2013]{nesvorny13} 
Nesvorn\'y, D.,  Kipping, D., Terrell, D., et al.  2013, \apj, 777, 3

\bibitem[2008]{oscoz08} 
Oscoz, A., Rebolo, R.,  L\'opez, R., et al. 2008, SPIE, 7014, 47

\bibitem[2017]{patra17} 
Patra, K. C., Winn, J. N., Holman, M. J., et al. 2017, \aj, 154, 4

\bibitem[2020]{patra20} 
Patra, K. C., Winn, J. N., Holman, M. J., et al. 2017, \aj, 159, 150

\bibitem[2002]{pepe02} 
Pepe, F., Mayor, M., Galland, F., et al. 2002, \aap, 388, 632

\bibitem[2008]{perruchot08}
Perruchot, S., Kohler, D., Bouchy, F., et al., 2008, 
SPIE, 7014, 70

\bibitem[2020]{petrucci20}
Petrucci, P. -O., Gronkiewicz, D., Rozanska, A., et al. 2020, \aap, 634, A85

\bibitem[2006]{pollacco06}
Pollacco, D. L., Skillen, I., Collier Cameron, A., et al. 2006, \pasp, 118, 140

\bibitem[2008]{pollacco08} 
Pollacco, D., Skillen, I., Collier Cameron, A., et al. 2008, \mnras, 385, 1576

\bibitem[1981]{popperetzel1981}
Popper, D. M., Etzel, P. B. 1981, \aj, 86, 102

\bibitem[1992]{press92}
Press, W. H., Teukolsky, S. A., Vetterling, W. T., et al. 
1992,  Numerical recipes in C. The art of scientiÞc 
computing, 2nd (Cambridge University Press) 

\bibitem[2001]{queloz01}
Queloz, D., Henry, G. W., Sivan, J. P., et al. 2001, \aap, 379, 279

\bibitem[2018]{rey18} 
Rey, J., Bouchy, F., Stalport, M., et al.\ 2018, \aap, 619, A115

\bibitem[2015]{ricker15} 
Ricker, G.R., Winn, J.N., Vanderspek, R., et al. 2015, JATIS, 1, 014003

\bibitem[2016]{santerne16} 
Santerne, A., Moutou, C.. Tsantaki, M., et al. 2016, \aap, 587, A64 

\bibitem[2006]{skrutskie06} 
Skrutskie, M. F.,  Cutri, R. M.,  Stiening, R., et al. 2006, \aj, 131, 1163

\bibitem[2011]{southworth2011} 
Southworth, J. 2011, \mnras, 417, 2166

\bibitem[2019]{southworth19} 
Southworth, J., Dominik, M., Jorgensen, U. G., et al. 2019, /mnras, 490, 4230

\bibitem[2016]{spake16} 
Spake, J. J.,  Brown, D. J. A., Doyle, A. P., et al. 2016, \pasp, 128, 024401

\bibitem[2017]{staab17} 
Staab, D., Haswell, C. A., Smith, G. D., et al. 2019, \mnras, 466, 738

\bibitem[2015]{tingley05} 
Tingley, B., Sackett, P. D. 2005, \apj, 627, 1011

\bibitem[2010]{torres10} 
Torres, G., Andersen, J., Gim\'enez, A. 2010, \aapr, 18, 67

\bibitem[2003]{udry03} 
Udry, S., Mayor, M., Santos, N. C. 2003n \aap, 407, 369

\bibitem[2017]{weiss17} 
Weiss, L. M., Deck, K. M., Sinukoff, E., et al. 2017, \apj, 153, 265

\bibitem[2009]{winn09} 
Winn, J. N., Johnson, J. A., Albrecht, S., et al. 2009, \apj, 703, L99

\bibitem[2013]{wittenmyer13} 
Wittenmyer, R. A., Wang, S., Horner, J., et al. 2013, \apjs, 208, 2

\bibitem[2013]{wu13} 
Wu, Y., Lithwick, Y.\ 2011, \apj, 772, 74 

\bibitem[2020]{yee20} 
Yee, S. W., Winn, J. N., Knutson, H. A., et al. 2020, \apj, 888, L5

\bibitem[2014]{xie14} 
Xie, J.-W. 2014, \apj, 210, 25

\end{thebibliography}
\end{document}